\documentclass[a4paper,aps,nofootinbib,amssymb,twocolumn,pra,superscriptaddress]{revtex4-2}
\usepackage{amsmath,amsfonts}
\usepackage{graphicx}
\usepackage{amssymb}
\usepackage{enumerate}
\usepackage{epsfig}
\usepackage{epstopdf}
\usepackage{mathrsfs}
\usepackage{amsthm}
\usepackage[T1]{fontenc}
\usepackage[latin9]{inputenc}
\usepackage[colorlinks,citecolor=blue,linkcolor=blue]{hyperref}
\usepackage{babel}
\newcommand{\E}{\mathcal{E}}

\newcommand{\ov}{\overline}

\newcommand{\ra}{\rangle}
\newcommand{\la}{\langle}

\newcommand{\be}{\begin{equation}}
\newcommand{\ee}{\end{equation}}

\newcommand{\D}{{\mathcal{D}}}

\newcommand{\ww}{\widetilde}
\newcommand{\dt}{\frac{d}{dt}}

\begin{document}

\title{ Preferred basis of states derived from eigenstate thermalization hypothesis
%\\ Preferred basis given by a Hamiltonian renormalized by a dissipative environment
%$\frac{1}{2} (m-1)(m+2)$ relations for steady states of open quantum systems with $m$ levels
}

\author{Hua Yan}
\affiliation{ Department of Modern Physics, University of Science and Technology of China,
 Hefei 230026, China}
\affiliation{CAS Key Laboratory of Microscale Magnetic Resonance,
University of Science and Technology of China, Hefei 230026, China}

\author{Jiaozi Wang}
\affiliation{ Department of Modern Physics, University of Science and Technology of China,
	Hefei 230026, China}
\affiliation{CAS Key Laboratory of Microscale Magnetic Resonance,
University of Science and Technology of China, Hefei 230026, China}
\affiliation{Department of Physics, University of Osnabr\"uck, D-49076 Osnabr\"uck, Germany}

\author{Wen-ge Wang}
\email{wgwang@ustc.edu.cn}
\affiliation{ Department of Modern Physics, University of Science and Technology of China,
	Hefei 230026, China}
\affiliation{CAS Key Laboratory of Microscale Magnetic Resonance,
University of Science and Technology of China, Hefei 230026, China}

% \date{\today}

\begin{abstract}
We study the long-time average of the reduced density matrix (RDM) of a central system
that is locally coupled to a large environment,  under a Schr\"{o}dinger evolution of the total system.
We consider a class of interaction Hamiltonian, whose environmental part
satisfies the so-called eigenstate thermalization hypothesis ansatz with a constant diagonal part
in the energy region concerned.
%On the eigenbasis of the central system's Hamiltonian,
%$\frac{1}{2} (m-1)(m+2)$
Relations among elements of the averaged RDM are derived.
When steady states of the central system exist,
these relations imply the existence of a preferred basis, which is given by the eigenbasis of
a renormalized self-Hamiltonian that
includes certain averaged impact of the system-environment interaction.
Numerical simulations performed for a qubit coupled to a defect Ising chain
confirm the analytical predictions.
\end{abstract}

\maketitle
%\tableofcontents

\section{Introduction}\label{sect-intro}

% \subsection{Motivation}
 Properties of small open quantum systems, which are coupled to large quantum environments,
 have attracted significant attention and been studied extensively in recent decades in various
 fields of physics \cite{leggett1987dynamics,breuer2002theory,alicki2007quantum,breuer2016colloquium,de2017dynamics}.
 Such a central system is described by its reduced density matrix (RDM) and may approach a steady
 state in many situations. For example,  it is now well known that  the phenomenon of decoherence,
 due to interactions with huge quantum environments,
 may happen in such a way that a RDM becomes approximately diagonal on a so-called preferred (pointer) basis
 of states (PBS)
 \cite{zurek1981pointer,zurek2003decoherence,schlosshauer2005decoherence,wiseman2009quantum,paz1999quantum,joos2013decoherence}.
 Under pure-dephasing interactions, decoherence has been studied well,
 with PBS given by eigenbases of self-Hamiltonians
 \cite{gorin2004connection,albash2015decoherence,weiss2012quantum}.
 However, under strong interactions and complex environments, with the self-Hamiltonians  negligible,
 PBS may be given by eigenbases of the interaction Hamiltonians
 \cite{zurek1981pointer,gorin2004connection}.

 The situation is much more complicated with a generic dissipative interaction,
 whose Hamiltonian is not commutable with the  central system's self-Hamiltonian,
 due to the interplay of decoherence and relaxation.
 In this generic case, knowledge about PBS is still far from being complete.
 Under a sufficiently weak  interaction and by a first-order
 perturbation theory, it was found that the system's eigenbasis is approximately a PBS
 under a quantum chaotic environment \cite{wang2008entanglement}.
 When the total system's eigenfunctions possess certain special randomness,
 a PBS (if existing) is given by the eigenbasis of a renormalized self-Hamiltonian
 \cite{he2014statistically}.
 These results are in agreement with a generic expectation for Markovian processes described by Lindblad master
 equations, as exemplified in solvable models \cite{breuer2002theory, alicki2007quantum,albash2015decoherence}.
 While, when non-Markovian effects due to dynamics of the total system are taken into account,
 nonnegligible off-diagonal elements of RDM have been found at long times in various spin-boson models
 on the spin's energy basis
 \cite{lee2012noncanonical,addis2014coherence,roszak2015decoherence,zhang2015role, guarnieri2018steady}.

 In this paper, we go further in the study of properties of steady states of small open systems,
 by directly computing their long-time averaged RDM under overall Schr\"{o}dinger evolutions of total systems.
 A key point of our approach is to consider those environments, for which
 the environmental parts of the interaction Hamiltonians satisfy
 the so-called eigenstate thermalization hypothesis (ETH) ansatz
 \cite{deutsch1991quantum,srednicki1994chaos,srednicki1999approach,rigol2008thermalization,
 d2016quantum,deutsch2018eigenstate}
 and their diagonal elements in the ansatz may be treated as constants within the energy regions of relevance.
 We are to derive $\frac 12 (m-1)(m+2)$ relations among elements of such an averaged RDM
 for a central system with a number of $m$ levels.
 When steady states exist, these relations imply that the central system should have a PBS,
 which is given by the eigenbasis of a renormalized self-Hamiltonian
 that includes certain impact of the system-environment interaction.

 The paper is organized as follows.
 In Sec.~\ref{sect-setup}, we specify the systems to be studied.
 In Sec.~\ref{sect-main-results}, we derive the above-mentioned relations.
 Some further discussions are given in Sec.\ref{sect-further-discussion}.
 Numerical simulations are presented in Sec.~\ref{sect-numerical},
 to illustrate validity of the analytical predictions for a qubit as the central system
 and a defect Ising chain as the environment.
 Finally, conclusions and discussions are given in Sec.~\ref{sect-conclusion}.

\section{Setup}\label{sect-setup}

 In this section, we discuss basic properties of the
 Hamiltonians of the systems to be studied.
 We use $S$ to denote the central system and use $\E$ to denote its (large) environment
 which consists of $N$ particles ($N\gg 1$).
 Hilbert spaces of $S$ and $\E$ are denoted by $\mathcal{H}^S$ and $\mathcal{H}^\E$,
 respectively, with dimensions $m$ and $d_\E$.
 The value of $m$ is required to be much smaller than the number of environmental levels
 that are of relevance effectively to the time evolution.

 The Hamiltonian of the total system is written as
\begin{equation}
 H = H^S + H^I + H^\E,
\end{equation}
 where $H^S$ and $H^\E$ are self-Hamiltonians of $S$ and $\E$, respectively,
 which are obtained in the weak coupling limit,
 and $H^I$ represents a local interaction Hamiltonian.
 Normalized eigenstates of the total system are denoted by $|n\ra$ with energies $E_{n}$
 in the increasing-energy order,
\begin{gather}\label{}
 H|n\rangle = E_{n}|n\rangle.
\end{gather}
 Normalized eigenstates of $H^S$ and of $H^\E$  are denoted by $|\alpha\rangle$ and $|i\rangle$, respectively,
 with labels $\alpha$ and $i$ as positive integers starting from $1$.
 The corresponding eigenenergies are denoted by $e^S_\alpha$ and $e_i$,
 respectively, both in the increasing-energy order,
\begin{subequations} \label{Sa}
\begin{align} \label{Sa-S}
 &H^S|\alpha\rangle = e^S_\alpha|\alpha\rangle,
 \\ &  H^\E|i\rangle = e_i|i\rangle,
\end{align}
\end{subequations}
 where for brevity we have omitted a superscript $\E$ for $e_i$.
 We use $\Delta_S$ to indicate the energy scope the central system $S$:
 \begin{gather}\label{Delta-S}
  \Delta_S := e^S_{m}- e^S_1.
 \end{gather}
%  and use $s_\E$ to indicate the mean nearest-level spacing of the environment in the studied energy region.
% For a small system, we consider values of $\Delta_S$ that are not large.
%  For a large environment,  since its density of states increases exponentially
%  with $N$, usually one has $\Delta_S \gg s_\E$, which is the case we consider here.

 We use $H^0$ to indicate the uncoupled Hamiltonian,
\begin{gather}\label{}
 H^0 = H^S + H^\E.
\end{gather}
 Its eigenstates are written as $|\alpha\rangle |i\rangle$, in short, $|\alpha i \rangle$, satisfying
 $H^0|\alpha i\ra = E_{\alpha i}|\alpha i\ra$, where $E_{\alpha i} = e^S_\alpha + e  _i$.
 % In the increasing-energy order, we write $|\alpha i\ra$ as $|E_r\ra$, with a one-to-one correspondence of $r\leftrightarrow (\alpha, i)$ and $ E_r = E_{\alpha i}$,
% \begin{gather}\label{}
%  H^0|E_r\ra = E_{r}|E_r\ra \qquad  (E_r \ge E_{r-1}).
% \end{gather}
The expansion of a state $|n\ra$ on the basis given by $|\alpha i\ra$ is written as
 \begin{equation}\label{|n>}
 |n\rangle = \sum_{\alpha i}C_{\alpha i}^n|\alpha i\rangle, %= \sum_{r}C_{r}^n|E_r\rangle,
\end{equation}
 with expansion coefficients $C_{\alpha i}^n$.
 % giving the energy eigenfunctions (EFs) of $|n\ra$ on the two bases.
 For the simplicity in discussion,
 we consider a product form of $H^I$,
 \footnote{Generalization to a generic local interaction Hamiltonian will be briefly discussed in Sec.~\ref{sect-g-int}.}
\begin{equation}\label{HI}
H^I =\lambda H^{IS}\otimes H^{I\E},
\end{equation}
 where $H^{IS}$ and $H^{I\E}$ are Hermitian operators acting on the two spaces of
 $\mathcal{H}^S$ and $\mathcal{H}^\E$, respectively, and $\lambda$ is
 a parameter for characterizing the coupling strength.
 Elements of  $H^{IS}$ and $H^{I\E}$ on  $|\alpha\ra$ and $|i\ra$ are written as
\begin{subequations}\label{HIS-alphabeta}
\begin{align}\label{}
 & H^{IS}_{\alpha \beta} = \la\alpha|H^{IS}|\beta\ra,
 \\ &  H^{I\E}_{ij} = \la i | H^{I\E}|j\ra.
\end{align}
\end{subequations}
 To describe locality of the interaction, we further divide the environment $\E$ into a small part denoted by $\E_1$
 and a large part denoted by $\E_2$,
 such that the system $S$ is coupled to $\E_1$ only.
 Then, $H^{I\E}$ is written as
\begin{gather}\label{HIE-local}
 H^{I\E} = H^{I\E_1} \otimes I^{\E_2},
\end{gather}
 where $H^{I\E_1}$ is an operator that acts on the Hilbert space of $\E_1$ and
 $I^{\E_2}$ indicates the identity operator on the Hilbert space of $\E_2$.
 \footnote{As a local operator, $H^{I\E_1}$ does not change with the environmental particle number $N$.}

 Although the exact condition under which  the ETH ansatz
 proposed in Ref.\cite{srednicki1999approach}  is applicable is still unclear,
 it is usually expected valid at least for local operators of many-body quantum chaotic systems
 \cite{d2016quantum,deutsch2018eigenstate}.
 Here, we assume that this ansatz is applicable to the operator $H^{I\E}$.
 According to this hypothesis,
 (1) diagonal elements $H^{I\E}_{ii}$ on average vary slowly with the eigenenergy $e_i$;
 (2) fluctuations of $H^{I\E}_{ii}$ possess certain random feature and are very small,
 scaling as $e^{-S(e)/2}$, where $S(e)$ is proportional to the particle number $N$ of $\E$
 and is related to the micro-canonical entropy in a semiclassical treatment;
 and (3) off-diagonal elements $H^{I\E}_{ij}$ with $i\ne j$ behave in a way similar to fluctuations of $H^{I\E}_{ii}$
 \cite{deutsch1991quantum,srednicki1994chaos,rigol2008thermalization,srednicki1999approach,
 d2016quantum, garrison2018does}.
 These predictions are written in the following concise form, usually
 referred to as the ETH ansatz,
 \begin{equation}\label{ETH}
 H^{I\E}_{ij} = {h}(e) \delta_{ij} + e^{-S(e)/2}g(e,\omega)R_{ij},
\end{equation}
 where  $e = (e_i+e_j)/2$, $\omega = e_j-e_i$, ${h}(e)$ is a slowly varying function of $e$,
 $g(e,\omega)$ is some smooth function,
 and $R_{ij}$ indicate random variables with a normal distribution (zero mean and unit variance).
 \footnote{Certain correlations among $R_{ij}$ have been observed
 numerically in some chaotic systems \cite{wang2022eigenstate},
 but, we do not discuss this possibility in this paper.}

 Analytical expressions of the functions $h(e)$ and $g(e,\omega)$ are still lacking.
 Numerically, three regimes have been observed for $|g(e,\omega)|$
 with respect to the order of per-site  energy  denoted by $\xi$,  provided that $e$ lies
 in the central region of the spectrum \cite{d2016quantum,de2017dynamics}.
 That is,  for $\omega\ll \xi$, it shows a plateau with a height proportional to $N^{1/2}$
 and a width  proportional to $N^{-2}$ \cite{rigol2008thermalization,khatami2013fluctuation};
 for large $\omega \gg \xi$, it decays exponentially;
 and,  for $\omega \sim \xi$, it is proportional to $\omega^{-1/2}$ in diffusive one-dimensional systems
 \cite{abanin2015exponentially,mukerjee2006statistical,brenes2020eigenstate,
 leblond2020universality}.

 For the simplicity in discussion, we set the initial  state of the total system at a time $t=0$
 as a pure state with a product form,
 \footnote{Discussions to be given below may be generalized, in a straightforward way, to a generic initial state
 written as $|\Psi(0)\rangle = \sum_\alpha c_{0\alpha}  |\alpha \rangle \otimes |\E_{0\alpha}\rangle$,
 if all the environmental states $|\E_{0\alpha}\rangle$ lie in the same energy shell $\Gamma^\E_0$ in Eq.(\ref{Gamma-E0}).
 }
 that is,
\begin{equation} \label{intialS}
 |\Psi(0)\rangle = |\phi_S\rangle \otimes |\E_0\rangle.
 \end{equation}
 Here, $|\phi_S\rangle$ indicates an arbitrary normalized state of the central system $S$, written as
\begin{gather}\label{phiS}
 |\phi_S\ra = \sum_\alpha c_{0\alpha} |\alpha\rangle;
\end{gather}
 and $|\E_0\ra$ is an arbitrary environmental state that lies within an energy shell
 denoted by $\Gamma^{\cal E}_0$,
\begin{equation}\label{E0}
 |\E_0\rangle = \sum_{e_i\in \Gamma^{\cal E}_0} c_{0i}|i\rangle.
\end{equation}
 The energy shell $\Gamma^{\cal E}_0$ is centered at an energy $e_0$ and has a width
 $\delta e_0$,
 \footnote{ See Sec.\ref{sect-main-condition} for a discussion about restriction to the width $\delta e_0$. }
 namely,
\begin{align}\label{Gamma-E0}
 \Gamma^{\cal E}_0 = [e_0-\delta e_0/2,e_0+\delta e_0/2 ].
\end{align}

% Expanded on the basis $|e_i\ra$, $|\E_0\ra$ is written as
%\begin{equation}\label{E0}
%|\E_0\rangle = \N_0 \sum_{e_i\in \Gamma^{\cal E}_0} c_{0i}|i\rangle.
%\end{equation}
% where the real and imaginary parts of $c_{0i}$ are independent Gaussian random
% variables, with mean zero and variance $1/2$, and $\N_0$ is the normalization factor.
% We use $N_{\Gamma_0^\E}$ to indicate the number of levels within the shell $\Gamma_0^\E$.
% Although being narrow, $\Gamma_0^\E$ is assumed to contain a large number of energy levels.
% It is seen that $\N_0^2 \simeq 1/N_{\Gamma_0^\E}$.

\section{Main result}\label{sect-main-results}

 In this section, we derive the main result of this paper,
 as relations among elements of the long-time averaged RDM.
 Specifically, we give some formal discussions in Sec.\ref{sect-inital-rdm},
 then,  in Sec.\ref{sect-delta-e}, derive an upper bound to the environmental energy region,
 which is of relevance effectively to the wave function at all times.
 The main result is derived in Sec.\ref{sect-main}
 and properties  of a main condition used in it are discussed in Sec.\ref{sect-main-condition}.

\subsection{Preliminary discussions}\label{sect-inital-rdm}

 The total system undergoes a Schr\"{o}dinger evolution,
\begin{gather}\label{Psi-t}
 |\Psi(t)\rangle  = e^{-iHt/\hbar}|\Psi(0)\rangle.
\end{gather}
 We write  $|\Psi(t)\ra$ in the following expansion with respect to the central system's states $|\alpha\ra$,
\begin{gather}\label{Psi-expan}
|\Psi(t)\rangle = \sum_{\alpha=1}^{m} |\alpha\rangle|\E_\alpha(t)\rangle,
\end{gather}
 and call $|\E_\alpha(t)\rangle$ the \emph{environmental branches}  of $|\Psi(t)\ra$.
 These branches, as vectors in the environmental Hilbert space, are written as
\begin{gather}\label{eq-branch}
 |\E_\alpha(t)\rangle = \la \alpha|\Psi(t)\rangle,
\end{gather}
 and are usually not normalized.
 Under the initial condition in Eq.(\ref{intialS}), it is direct to find that
\begin{equation}
	\label{branch-eq}
	|\E_\alpha(t)\rangle=\sum_{\beta}c_{0\beta}\langle \alpha|e^{-iHt/\hbar}
|\beta\rangle|\E_0\rangle,
\end{equation}
 and
\begin{equation} \label{alpha-t}
 i\hbar\dt |\E_\alpha(t)\rangle = H_{\alpha\alpha}|\E_\alpha(t)\rangle
 + \sum_{\beta \ne \alpha} H_{\alpha\beta}|\E_\beta(t)\rangle,
\end{equation}
 where $H_{\alpha\beta}$ indicate operators that act on the Hilbert space
 of the environment, as defined below,
\begin{gather}\label{H-subspace}
H_{\alpha\beta} : =\langle \alpha|H|\beta\rangle.
\end{gather}

 By definition, the RDM of the system $S$, denoted by $\rho^S(t)$,
 is given by $\rho^S(t) = \text{Tr}_\E \rho(t)$, where $\rho(t) = |\Psi(t)\rangle\langle\Psi(t)|$.
 It is easy to check that elements of the RDM on the basis $\{|\alpha\ra \}$, namely
 $\rho_{\alpha\beta}^S(t) = \langle \alpha|\rho^S(t)|\beta\rangle$,
 have the following expression,
\begin{equation}\label{rho-abt}
\rho_{\alpha\beta}^S(t) = \langle \E_\beta(t)|\E_\alpha(t)\rangle.
\end{equation}
 Making use of Eq.(\ref{alpha-t}),  after some deviation,
 one finds that the elements $\rho_{\alpha\beta}^S(t)$ satisfy the following equation
 (see Appendix \ref{append1}),
\begin{eqnarray}\label{drhot}
i\hbar \frac{d\rho_{\alpha\beta}^S(t)}{dt}
%&=& i\langle\E_\beta(t)|\overleftarrow{\partial}_t|\E_\alpha(t)\rangle + i\langle\E_\beta(t)|
%\overrightarrow{\partial}_t|\E_\alpha(t)\rangle\nonumber\\
= W_{\alpha\beta}^{(1)} + \lambda W_{\alpha\beta}^{(2)},
\end{eqnarray}
 where
\begin{subequations}\label{W1-2}
\begin{eqnarray}\label{w1}
 W_{\alpha\beta}^{(1)} &=& (e_\alpha^S-e_\beta^S)\rho_{\alpha\beta}^S(t),
% =\langle \alpha|[H^S,\rho^S(t)]|\beta\ra,
 \\ \label{w2}
 W_{\alpha\beta}^{(2)} &=& \sum_{\gamma=1}^{m} H_{\alpha \gamma}^{IS}
 F_{\beta \gamma}(t)-\sum_{\gamma=1}^{m} H^{IS}_{\gamma\beta}
 F_{\gamma\alpha}(t).
\end{eqnarray}
\end{subequations}
 Here, $F_{\alpha\beta}(t)$ indicate $c$-number quantities defined below,
\begin{equation}\label{HIE-ab}
F_{\alpha\beta}(t)  :=\langle \E_\alpha(t)|H^{I\E}|\E_\beta(t)\rangle,
\end{equation}
 and, from them, we define the following operator,
\begin{align} \label{F(t)}
	F(t) :=\sum_{\alpha\beta}F_{\alpha\beta}(t)|\alpha\rangle\langle \beta|.
\end{align}
 It is easy to check that $W_{\alpha\beta}^{(1)}$ and $W_{\alpha\beta}^{(2)}$ have the following concise expressions,
\begin{subequations}\label{w12-conc}
\begin{align}\label{}
 & W_{\alpha\beta}^{(1)} =\langle \alpha|[H^S,\rho^S(t)]|\beta\ra, \label{w1-conc}
 \\ & W_{\alpha\beta}^{(2)} =\langle \alpha|[H^{IS}, F^T(t)]|\beta\ra, \label{w2-conc}
\end{align}
\end{subequations}
% (\textbf{check W2})
 where ${F}^{T}$ indicates the transposition operator of ${F}$, which is defined on the eigenbasis of $H^S$.

 We use an overline to indicate the long-time average of a term.
 For example, the long-time average of the RDM is written as $\ov{\rho}^S$,
\begin{equation}\label{ov-rho}
\ov{\rho}^S   =\lim_{t\to\infty} \frac{1}{t}\int_0^t \rho^S(t')dt'.
\end{equation}
 Clearly, in the case that a steady state of the RDM exists, it is given by $\ov{\rho}^S$.
 Since the elements $\rho_{\alpha\beta}^S(t)$ have bounded values,
 the long-time average of $d\rho_{\alpha\beta}^S(t)/{dt}$ must be zero, i.e.,
 $\ov{{d\rho_{\alpha\beta}^S(t)}/{dt}}=0$.
 Then, Eq.(\ref{drhot}) gives that
\begin{gather}\label{ov-WW=0}
 \ov {W}_{\alpha\beta}^{(1)} + \lambda \ov{W}_{\alpha\beta}^{(2)} =0.
\end{gather}
Substituting the explicit expressions of ${W}_{\alpha\beta}^{(1)}$
 and ${W}_{\alpha\beta}^{(2)}$ in Eq.(\ref{W1-2}) into Eq.(\ref{ov-WW=0}), one finds
 the following formal relation for the long-time averaged RDM:
\begin{align}\label{rho-S-0}
	(e_\alpha^S-e_\beta^S)\ov{\rho}_{\alpha\beta}^S +\lambda \sum_{\gamma=1}^{m}
\Big[H_{\alpha \gamma}^{IS} \ov{ F}_{\beta \gamma}-H^{IS}_{\gamma\beta}\ov{ F}_{\gamma\alpha} \Big] =0.
\end{align}	
 It is straightforward to check that a concise form of Eq.(\ref{rho-S-0}) is written as
\begin{align}
	\label{rho-S-1}
	[H^S,\ov{\rho}^S]+\lambda[H^{IS}, \ov{F}^{T}]=0.
\end{align}
% Note that, as a transposition operator, ${F}^{T}$ is basis-dependent.
%  In the case of a TLS as the central system, Eq.(\ref{rho-S-0}) has the following simple form,
% \begin{gather}\label{tls-rho}
% \overline{\rho}_{\alpha\beta}^S  = \eta_d \ov{F}_{\beta \alpha}+\eta_r  (\ov{F}_{\beta\beta}- \ov{F}_{\alpha\alpha})
%  \ \quad \  (\alpha \ne \beta),
% \end{gather}
%  where
% \begin{gather}\label{eta-dr}
% \eta_{d} :=\frac{H_{\alpha\alpha}^{IS}-H_{\beta\beta}^{IS}}{e_{\beta}^{S}-e_{\alpha}^{S}},
% \quad\eta_{r} :=\frac{H_{\alpha\beta}^{IS}}{e_{\beta}^{S}-e_{\alpha}^{S}}.
% \end{gather}
%  The quantity $\eta_d$ gives a measure to the relative strength of dephasing,
%  while, $\eta_r$ for the relative strength of relaxation or dissipation.
%  (The subscript ``$d$'' in $\eta_d$ stands for ``dephasing'' and ``$r$'' in $\eta_r$ for ``relaxation''.)

\subsection{Effective environmental energy region}\label{sect-delta-e}

 In this section, we discuss an environmental energy region,
 within which all the branches $|\E_\alpha(t)\rangle$ effectively lie for all the times $t$, and indicate it by $\Gamma^\E$.
 We do not need to find the smallest one of this type of region.
 Instead, we consider a region that has the following simple form,
 \begin{align}
	\Gamma^\E =[e_0-\delta e/2,e_0+\delta e/2],
\end{align}
 centered at the initial center $e_0$ and with a width $\delta e$.

Below, we derive an expression for $\delta e$, as an upper bound to the width of the energy
 region that effectively contains all $|\E_\alpha(t)\rangle$.
 For this purpose, we need to analyze the components $\la i|\E_\alpha(t)\rangle$,
\begin{align}\label{Eat-expan-1}
	\la i|\E_\alpha(t)\rangle = \sum_{\beta,n} \sum_{e_j\in\Gamma_0^\E}
 c_{0\beta}c_{0j}C^{n*}_{\beta j}C^n_{\alpha i}e^{-iE_nt},
\end{align}
 which is directly obtained by making use of Eqs.(\ref{intialS})-(\ref{E0}) and (\ref{Psi-t})-(\ref{eq-branch}).
 Initially, with $e^{-iE_nt}=1$ at $t=0$,
 due to correlations among the terms of $(C^{n*}_{\beta j}C^n_{\alpha i})$ of different indices $n$,
 which originate from the completeness of the states $|n\ra$ as a basis in the total Hilbert space,
 nonzero values of the rhs of Eq.(\ref{Eat-expan-1}) are restricted within the initial energy region $\Gamma^\E_0$.
 With increase of the time $t$, the phases $e^{-iE_nt}$ gradually destroy the above-mentioned correlations
 and, as a result, the energy region that is effectively occupied by $|\E_\alpha(t)\rangle$ expands.
 Cutting all the correlations by taking an absolute value for each summed term on the rhs of  Eq.(\ref{Eat-expan-1}),
 one gets an upper bound to $ |\la i|\E_\alpha(t)\rangle|$:
\begin{align}\label{Eat-expan-abs}
	|\la i|\E_\alpha(t)\rangle| \le \sum_{e_j\in\Gamma_0^\E}
\sum_{\beta,n} \left| c_{0\beta}c_{0j}C^{n*}_{\beta j}C^n_{\alpha i} \right|.
\end{align}

 When using the rhs of Eq.(\ref{Eat-expan-abs}) to get an upper bound
 to the environmental energy region that effectively contains
 $|\E_\alpha(t)\rangle$, exact values of the nonzero coefficients $c_{0\beta}$ and $c_{0j}$ are not important.
 Hence, we may focus on the values of $C^{n*}_{\beta j}C^n_{\alpha i}$.
 In particular, for the eigenfunction (EF) $C^n_{\alpha i}$ of each state $|n\ra$,
 what is of relevance is its main-body region,
 within which the main population lie up to a small error indicated by $\epsilon$.
 Energetically, such a main-body region consists of those uncoupled states $|\alpha i\ra$, whose energies $E_{\alpha i}$
 are around the exact energy $E_n$ within a scope which we indicate by $w^\epsilon_{n}$.
 More exactly, the set of the indices of these uncoupled states, indicated by $\Omega^\epsilon_n$,
 is written as
\begin{align}\label{}
 \Omega^\epsilon_n = \left\{ (\alpha,i):  |E_{\alpha i} - E_n| \le \frac 12 w^\epsilon_{n} \right\}.
\end{align}
 Then, the main-body region of the EF $C^n_{\alpha i}$  satisfies the following requirement,
\begin{align}\label{width-eq}
 \sum_{(\alpha, i) \in \Omega^\epsilon_n} \left| C^n_{\alpha i} \right|^2 \doteq 1- \epsilon  \quad (\epsilon\ll 1),
\end{align}
 where ``$\doteq$'' means that the set $\Omega^\epsilon_n$
 is chosen such that the left hand side of Eq.(\ref{width-eq}) is the closest to its right hand side.

 \begin{figure}
	 \includegraphics[width=1.0\linewidth]{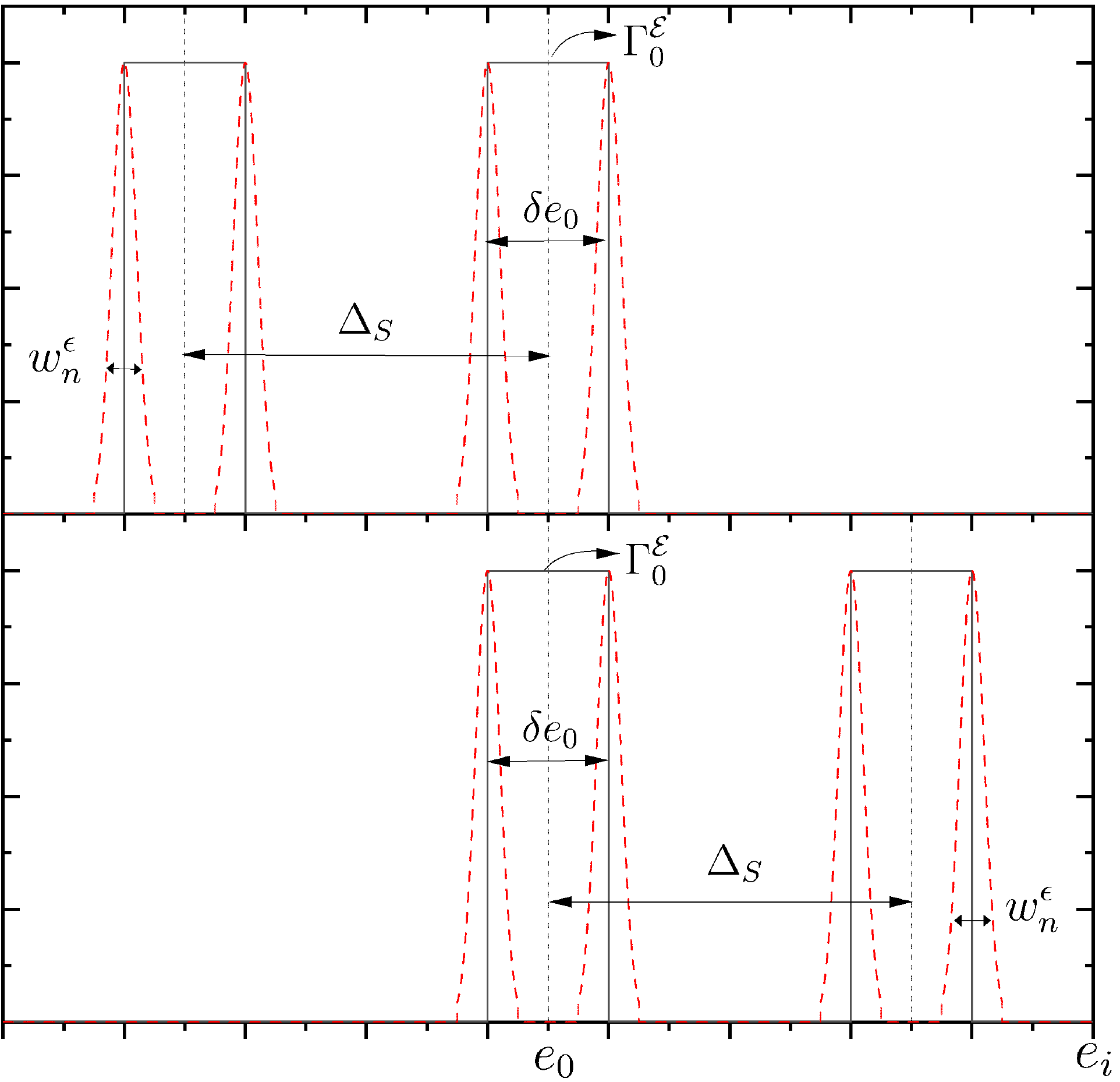}
	 \caption{A schematic illustration for the energy region $\Gamma^\E$.
 Upper panel: the environmental branch that moves to the farthest left from the initial shell $\Gamma^\E_0$
 due to the system-environment interaction.
 Lower panel: the branch that moves to the farthest right.
}
	 \label{fig:scheme}
 \end{figure}

 For a product $C^{n*}_{\beta j}C^n_{\alpha i}$ to give a nonnegligible contribution
 to the rhs of Eq.(\ref{Eat-expan-abs}),
 both of the two basis states $|\beta j\ra$ and $|\alpha i\ra$ should lie in the main-body region of $|n\ra$.
 As as result, the expansion from $\Gamma^\E_0$ to $\Gamma^\E$ should be influenced mainly by
 two factors: widths of main-body regions of the EFs and the central system's energy differences.
 We use $w^\epsilon_{\rm max}$ to indicate the maximum value of $w^\epsilon_{n}$
 for those states $|n\ra$ that are of relevance to the time evolution of the initial state.
 Then, noting that $\Delta_S$ in Eq.(\ref{Delta-S}) gives the maximum of $|e^S_\alpha - e^S_\beta|$,
 we find the following expression of $\delta e$,
\begin{align}\label{delta-e}
 \delta e = \delta e_0+2\Delta_S+w^\epsilon_{\rm max},
\end{align}
 as illustrated in Fig.\ref{fig:scheme}.

 Two remarks are in order: (1) At $\lambda=0$, the real width is just $\delta e_0$,  smaller than $\delta e $.
  (2) When the states $|\alpha i\ra$ are sufficiently coupled by the interaction,
 it is possible for $\delta e$ in Eq.(\ref{delta-e}) to be close to the width of the energy region
 that is really occupied.

 \subsection{Relations among elements of averaged RDM}\label{sect-main}

 In this section, we derive the main result, as relations that the elements of $\ov \rho^S$ satisfy.
 To this end, let us expand the environmental branches $|\E_\alpha(t)\rangle$ as,
\begin{gather}	\label{fai}
	|\E_\alpha(t)\rangle = \sum_i f_{\alpha i }(t)|i\rangle
\end{gather}
 with expansion coefficients $f_{\alpha i }(t)$.
 Substituting Eq.\eqref{fai} into Eqs.\eqref{rho-abt} and \eqref{HIE-ab},
 taking the long-time average, and making use of the fact that
 all the environmental branches effectively lie within the energy region $\Gamma^\E$,
 one finds the following expressions of $\overline{\rho}_{\alpha\beta}^S$
  and $\overline{F}_{\beta\alpha}$,
\begin{align}\label{}
 \label{rho-lta}
 &  \ov{\rho}_{\alpha\beta}^S = \sum_i \overline{f_{\beta i}^*f_{\alpha i }}\simeq \sum_{e_i\in\Gamma^\E}
  \overline{f_{\beta i}^*f_{\alpha i }},
  \\ & \ov{F}_{\beta\alpha} = \sum_{i,j}\overline{f_{\beta j}^*
  f_{\alpha i}}H^{I\E}_{ji}\simeq \sum_{e_i,e_j\in\Gamma^\E}\overline{f_{\beta j}^*f_{\alpha i}}H^{I\E}_{ji}.
  \label{f-lta}
\end{align}
 Substituting the ETH ansatz \eqref{ETH} into Eq.\eqref{f-lta}, one gets that
  \begin{align}\label{Fab-1}
	\overline{F}_{\beta\alpha} \simeq \sum_{e_i\in {\Gamma}^\E}h(e_i)\overline {f_{\beta i}^*f_{\alpha i}}
 + \Delta_{\alpha\beta},
  \end{align}
  where $ \Delta_{\alpha\beta}$ indicates a fluctuation term, given by
\begin{align}\label{Delta-ab}
	  \Delta_{\alpha\beta} =\sum_{i,j}
\overline {f_{\beta i}^*f_{\alpha j}}e^{-S(e)/2}g(e,\omega)R_{ij}.
  \end{align}
% Clearly, the summation on the rhs of Eq.(\ref{Delta-ab}) is effective for $e_i,e_j\in \Gamma^\E$.

 Generically, the two operators $\ov F$ and $\ov \rho^S$ do not have a simple relationship.
 One key observation made here is that they may possess a simple relationship, if
 the function $h(e)$ is approximately a nonzero constant  within the
 energy shell $\Gamma^\E$.
 More exactly, the condition is that
 \begin{align}\label{cond-h}
	\left| \frac{1}{h_0} \Delta h \right| \le \epsilon_h \quad \text{with $\epsilon_h \ll 1$},
\end{align}
 where $\epsilon_h$ is a parameter much smaller than $1$,
 $h_0 \equiv h(e_0)$ with $h_0 \ne 0$ (even in the limit of $N \to \infty$),
 and $\Delta h$ indicates the maximum difference between $h_0$ and $h(e_i)$ within $\Gamma^\E$:
\begin{align}\label{Delta-h}
  \Delta h = \max_{e_i\in{\Gamma}^\E}|h(e_i)-h_0|.
\end{align}

 To show the above-mentioned relationship,
 we note that, when the condition in Eq.(\ref{cond-h}) is satisfied, Eqs.(\ref{rho-lta})
 and (\ref{Fab-1}) imply that
\begin{align}\label{Fab-2}
 \overline{F}_{\beta\alpha} \simeq h_0\overline{\rho}^S_{\alpha\beta}+\Delta_{\alpha\beta},
\end{align}
 or, equivalently,
\begin{equation}	\label{eq:ov-F}
	\ov{F}^{T}\simeq h_0\ov{\rho}^S+\Delta,
\end{equation}
 where $\Delta$ is a fluctuation operator,
\begin{align}\label{eq:Delta}
\Delta :=\sum_{\alpha,\beta}\Delta_{\alpha\beta}|\alpha\rangle\langle\beta|.
\end{align}

 From the rhs of Eq.(\ref{Delta-ab}), one sees three factors that influence the $N$-dependence of $\Delta$.
 The first one is the exponential decay of $e^{-S(e)/2}$, with $S(e) \sim N$.
 The second factor is given by the unknown ETH-ansatz function $g(e,\omega)$,
 for which numerical simulations show a polynomial increase of $N^\gamma$, with $\gamma=1/2$
 in some (diffusive one-dimensional) systems
 \cite{luitz2016anomalous,d2016quantum}.
 The third factor  lies in the summation over the indices $i$ and $j$ and
 the long-time average term $\overline {f_{\beta i}^*f_{\alpha j}}$.
 As shown in Appendix \ref{append3:norm-bound-eth}, due to the randomness of $R_{ij}$,
 contribution from the third factor is negligible compared with the first factor.
 (See Eq.(\ref{norm-Delta-ub}) for an upper bound to the norm of $\Delta $.)
 Therefore, the $N$-scaling behavior of the fluctuation operator $ \Delta $
 is dominated by the exponential decay term $e^{-S(e)/2}$.
 Due to this exponential decay, as well as
 the fact that $h_0 \ne 0$ in the limit of large $N$ and that each RDM has a unit trace,
  one gets that
\begin{align}\label{}
 \| \Delta \| \ll \| h_0 \ov{\rho}^S \| \quad \text{at sufficiently large $N$}.
\end{align}
 Then, Eq.\eqref{eq:ov-F} gives the following  relation between the two operators of $\ov F$ and $\ov\rho$:
\begin{align}\label{F-rho-relation}
 \ov{F}^{T}\simeq h_0\ov{\rho}^S \quad \text{at sufficiently large $N$.}
\end{align}

 The main result of this paper is obtained by substituting Eq.(\ref{F-rho-relation}) into Eq.\eqref{rho-S-1},
\begin{align}
 \label{H-rho-comm}
	[\widetilde{H}^S,\ov{\rho}^S]\simeq 0,
\end{align}
 which holds under the condition of Eq.(\ref{cond-h}) and at sufficiently large $N$.
 Here, $\widetilde{H}^S$ is a renormalized self-Hamiltonian of the central system, defined by
\begin{align}
	\label{Hs-normal}
	\widetilde{H}^S := H^S + \lambda h_0H^{IS},
\end{align}
 which includes certain averaged impact of the system-environment interaction.
 From Eq.(\ref{H-rho-comm}), one sees that, if a PBS exists, it should be
 given by the eigenbasis of the renormalized self-Hamiltonian $\widetilde{H}^S$.
 Writing Eq.\eqref{H-rho-comm} explicitly, one gets that
\begin{align} \label{relation-explicit}
 (e_\alpha^S-e_\beta^S)\overline{\rho}^S_{\alpha\beta}+\lambda h_0\sum_\gamma[H^{IS}_{\alpha\gamma}
 \ov{\rho}^S_{\gamma\beta}-\ov{\rho}^S_{\alpha\gamma}	H^{IS}_{\gamma\beta}]\simeq 0.
\end{align}
 This gives $m(m-1)/2$ relations among elements of the averaged RDM
 for $\alpha\ne\beta$ and $(m-1)$ relations for $\alpha=\beta$.

 As an illustration of the above result, let us consider a nondegenerate two-level system (TLS), with $e_{2}^{S} \ne e_{1}^{S}$.
 From Eq.(\ref{relation-explicit})  with $\alpha \ne \beta$, one gets that
\begin{align}
	\label{eq:tls-rhos}
	\overline{\rho}^S_{12}\simeq\frac{\lambda\eta_r h_0}{1-\lambda\eta_d h_0}
(\overline{\rho}^S_{22}-\overline{\rho}^S_{11}),
\end{align}
 where
 \begin{gather}\label{eta-dr}
	\eta_{d} =\frac{H_{11}^{IS}-H_{22}^{IS}}{e_{2}^{S}-e_{1}^{S}},
	\quad\eta_{r} =\frac{H_{12}^{IS}}{e_{2}^{S}-e_{1}^{S}}.
	\end{gather}
	 The quantity $\eta_d$ gives a relative measure for the strength of dephasing,
	 while, $\eta_r$ gives a relative measure for the strength of relaxation (dissipation).
 Meanwhile, in the case of $\alpha=\beta$, one gets that
	 \begin{equation}
		 H^{IS}_{12}\ov{\rho}^S_{21}- \ov{\rho}^S_{12}H^{IS}_{21} \simeq 0,
	 \end{equation}
 which implies approximate realness of  the product $H^{IS}_{12}\ov{\rho}^S_{21}$.

\subsection{$N$- and $\lambda$-relevance to the condition (\ref{cond-h})}\label{sect-main-condition}

 In this section, we discuss relevance of the particle number $N$ to Eq.(\ref{cond-h}),
 a main prerequisite for the above-derived main result, as well as relevance of the interaction strength $\lambda$.
 \footnote{
 One may note that Eq.(\ref{cond-h}) is always satisfied,
 in the case that EFs of the quantum chaotic environment
 may be  effectively described by the random matrix theory (RMT).
 In fact, in this case, $h(e)$ is  a constant, given by $h(e) = \text{tr}(H^{I\E})/d_\E$ \cite{d2016quantum}.
}
 Basically, Eq.(\ref{cond-h}) requires that the environmental energy shell $\Gamma^\E$
 should be ``sufficiently narrow'', such that the function $h(e)$ may be approximately taken as a constant within it,
 compared with its nonzero central value $h_0$.
 Below, we give a detailed discussion of the exact meaning of ``being sufficiently narrow''.

\subsubsection{Relevance of the particle number N}

 Relevance of $N$ to Eq.(\ref{cond-h}) comes mainly from two aspects: the width $\delta e$ of $\Gamma^\E$
 in Eq.(\ref{delta-e}) and the ETH-ansatz function $h(e)$.
 The width $\delta e =\delta e_0+2\Delta_S+w^\epsilon_{\rm max}$ contains three terms.
 Clearly, $\Delta_S$,  the central system's energy scope, is $N$-independent.
 The $N$-dependence of $\delta e_0$ is usually determined according to the problem at hand,
 particularly, to quantities of final interest;
 e.g., it may be taken as a constant, or as some polynomial function of $N$.

 The situation with $w^\epsilon_{\rm max}$,  the maximum width of relevant EFs of the total system
 on the uncoupled energy basis,  is more complicated.
 In fact, presently, still not much is known analytically about widths of the EFs.
 It seems reasonable to assume that $w^\epsilon_{\rm max} \sim N^\mu$ with some parameter
 $\mu $ the value of which may be model-dependent.
 By a first-order perturbation-theory treatment to long tails of EFs in certain model,
 it was found that $\mu <0$ \cite{wang2012statistical};
 while, a study of higher-order contributions is still under investigation \cite{EF-semp}
 by making use of a semiperturbative theory \cite{wang2019convergent,wang1998structure,wang2000perturbative,wang2002nonperturbative}.

 The ETH ansatz does not assume any specific form of the function $h(e)$.
 According to numerical simulations with the help of some analytical analysis
\cite{kim2014testing,brenes2020eigenstate, d2016quantum,brenes2020low},
 $h(e)$ was found approximately a function of  per-site energy,
\begin{align}\label{he-norma}
	h(e_i) \approx \widetilde{h}(e_i/N),
\end{align}
 where $\widetilde{h}(x)$ is some smooth function of $x$, independent of $N$.
 Then, Taylor's expansion gives that
\begin{align}
   \label{eq:eth-diag-diff}
	h(e_i)-h(e_j) =  \ww h'(e_i/N)  \frac{e_i-e_j}{N} + O_2(\frac{e_i-e_j}{N}),
\end{align}
 where $\ww h'(x)$ indicates the derivative of $\ww h(x)$ and
 $O_2$ represents the second and higher order terms of the expansion.

 To be specific, let us discuss a case,
 in which the initial width $\delta e_0$  increases slower than $N$ such that
\begin{align}\label{de0/N-Nlimit}
 \lim_{N \to \infty} \frac{\delta e_0}{N} =0.
\end{align}
 This case may be met quite often practically.
 Note that Eq.(\ref{de0/N-Nlimit}) does not really require narrowness of the initial shell $\Gamma^\E_0$;
 e.g., it holds for $\delta e_0 \sim N^b$ with a parameter $b <1$.
 Then, as long as $\mu <1$ for $w^\epsilon_{\rm max} \sim N^\mu$,
 Eq.(\ref{de0/N-Nlimit}) implies that
\begin{align}\label{de/N-Nlimit}
 \lim_{N \to \infty} \frac{\delta e}{N} =0.
\end{align}
 This implies that the ratio $|e_i-e_j|/{N}$ should approach zero in the limit of large $N$
 for $e_i, e_j \in \Gamma^\E$.
 If $\ww h' \ne 0$, then, according to Eq.(\ref{eq:eth-diag-diff}),
 the difference $[h(e_i)-h(e_j)]$ is approximately given by the first-order term  at sufficiently large $N$.
 As a consequence, $h(e)$ is approximately a linear function
 within $\Gamma^\E$ and $\Delta h$ in Eq.(\ref{Delta-h}) is written as
\begin{align}\label{eq:fluc-h-estimate}
 \Delta h \approx \frac{\delta e}{2N} \left| \ww h' \left( \frac{e_0}N \right) \right| .
\end{align}
 One sees that, as long as $\left| \ww h' \left( {e_0}/N \right) \right|$ has a finite upper bound,
 $\Delta h/h_0 \to 0$ in the limit of $N\to \infty$.
 Otherwise, i.e., if  $\ww h' = 0$, one may consider the second-order term (if nonzero)  in Eq.(\ref{eq:eth-diag-diff})
 and, following arguments similar to those given above, reach the same conclusion.
 Similar arguments also apply, when higher-order terms dominate.
 Therefore, for systems with $\mu <1$, under an initial condition satisfying Eq.(\ref{de0/N-Nlimit}),
 the condition (\ref{cond-h}) is usually fulfilled at sufficiently large $N$.

\subsubsection{Relevance of the interaction strength}\label{sect-relevance-IS}

 Among the three terms of $\Delta_S$, $\delta e_0$,  and $w^\epsilon_{\rm max}$ in $\delta e$,
 only the EF width $w^\epsilon_{\rm max}$ depends on  the interaction strength $\lambda$.
 As is well known, usually, $w^\epsilon_{\rm max}$ increases with increasing $\lambda$,
 when other parameters in the total Hamiltonian are fixed.
% This is in contrast to the above-discussed  decreasing behavior of $w^\epsilon_{\rm max}$
% with increasing $N$.
 It is reasonable to expect that dependence of $w^\epsilon_{\rm max}$  on the pair of $(N,\lambda)$
 may behave in a quite complicated way.
 A full understanding of this behavior is beyond the scope of this investigation.
 Below, for the sake of clearness in discussion,
 we usually consider a fixed  value of $N$ when discussing influence of $\lambda$.

 To study influence of the interaction strength $\lambda$ on the condition in Eq.(\ref{cond-h}),
 let us consider a case in which Eq.(\ref{cond-h}) is satisfied at $\lambda=0$ with $w^\epsilon_{\rm max} =0$.
 For example, one has such a case, if the initial shell $\Gamma^\E_0$ is sufficiently narrow and the value of
 $\Delta_S$ is sufficiently small.
 With increase of $\lambda$ from $0$,  the value of $\delta e$ increases due to the increase of $w^\epsilon_{\rm max}$.
 At a small $\lambda$, the width $w^\epsilon_{\rm max}$ is still small and, as a result,
 Eq.(\ref{cond-h}) is also satisfied.

 When the value of $\lambda$ increases beyond some regime,
 usually, it is possible for $\Delta h$ to become
 sufficiently large such that Eq.(\ref{cond-h}) gradually becomes invalid.
 Note that the width $w^\epsilon_{\rm max}$ has no upper bound,
 because it should increase (approximately) linearly with $\lambda$
 when the interaction Hamiltonian dominates in the total Hamiltonian.
 To be quantitative, related to breakdown of Eq.(\ref{cond-h}),
 one may consider a value of $\lambda$, indicated as $\lambda_h$,
 at which the value of $|\Delta h / h_0|$ first reaches $\epsilon_h$ when $\lambda$ increases from $0$.
 Making use of Eqs.(\ref{eq:fluc-h-estimate}) and (\ref{delta-e}), from Eq.(\ref{cond-h}) one gets that
\begin{align}\label{rela-lamb-h}
	\delta e_0+2\Delta_S+w^\epsilon_{\rm max}(\lambda_h)
 \approx 2N \epsilon_h  \left| \frac{h_0}{\ww h' \left( \frac{e_0}N \right)} \right|.
\end{align}
 Two properties are seen from Eq.(\ref{rela-lamb-h}):
 (1) Since the width $w^\epsilon_{\rm max}$ usually increases with increasing $\lambda$,
 for systems with $\mu<1$,  the value of $\lambda_h$ may increase with increasing $N$;
 and, (2) $\lambda_h$ should increase with decreasing $\Delta_S$,
 if other parameters are fixed.

\section{Further discussions}\label{sect-further-discussion}

 In this section, we discuss two situations, in which
 some modified versions of the RDM relations given in the main result still hold when
 some restrictions used above are loosened.
 In Sec.\ref{sect-weak-coupling}, we derive RDM relations  in the weak coupling limit,
 without the restriction of Eq.(\ref{cond-h}).
 In Sec.\ref{sect-g-int}, we show that the main result may be generalized to
 a generic local interaction Hamiltonian.

\subsection{Offdiagonal elements at very weak couplings}\label{sect-weak-coupling}

 In this section, in the weak coupling limit of $\lambda$, without using the condition in Eq.(\ref{cond-h}),
 we derive an expression for offdiagonal elements of the averaged RDM of nondegenerate levels,
 by employing a first-order perturbation treatment.
 In this limit, diagonal elements of RDM keep approximately constants, directly given by the initial condition:
\begin{align}\label{rho-aa=c0a}
 \overline{\rho}^S_{\alpha \alpha} \simeq |c_{0\alpha}|^2 \qquad \text{for nondegenerate levels $\alpha$}.
\end{align}

 To be specific, below, we consider two arbitrary nondegenerate levels of the central system $S$,
 indicated by $\alpha$ and $\beta$ with $e_\beta^S \ne  e_\alpha^S$.
% Let us first consider the zeroth-order terms.
 The zeroth-order branches, denoted by $|\E^{\rm 0th}_\alpha(t)\ra$,
 are computed by the Schr\"{o}dinger evolution of the initial state $|\Psi(0)\rangle$ under
 the uncoupled Hamiltonian $H^0$.
 Noting Eq.(\ref{intialS}), one directly gets that
\begin{align}\label{}\notag
 |\E^{\rm 0th}_\alpha(t)\ra &  = \la \alpha|e^{-iH^0t} |\Psi(0)\ra
 \\ & = c_{0\alpha} e^{-ie^S_\alpha t}  \sum_{e_j\in \Gamma^{\cal E}_0} e^{-ie_j t} c_{0j}|j\rangle.
\label{E0th}
\end{align}
 Substituting Eq.(\ref{E0th}) into Eq.(\ref{rho-abt}) and noting that $e_\beta^S \ne  e_\alpha^S$,
 one sees that $\ov{\rho}_{\alpha\beta}^S$ has a vanishing zeroth-order term.

 The zeroth-order term of $\overline{F}_{\alpha\beta}$, indicated as $\overline{F}_{\alpha\beta}^{\rm 0th}$,
 is computed by substituting Eq.(\ref{E0th}) into Eq.\eqref{HIE-ab} and taking the long-time average.
 Noting that the chaotic environment $\E$ has a nondegenerate spectrum,
 direct computation gives that
 \footnote{For an environment that possesses a degenerate spectrum,
 one may divide the set of those labels $i$, for which $e_i\in \Gamma^\E_0$,
 into subsets according to the degeneracy.
 We denote the subsets by $\D_q$ with a label $q$, such that $e_i=e_j$ for all $i,j\in \D_q$.
 Then, it is easy to find that
\begin{align}\label{h1-2}
 h_1 = \sum_{q } \sum_{i,j \in \D_q} c_{0i}^* H^{I\E}_{ij} c_{0j}.
\end{align}
}
\begin{align}\label{F0th-ab}
\overline{F}_{\alpha\beta}^{\rm 0th}=|c_{0\alpha}|^2  h_1 \delta_{\alpha \beta},
\end{align}
 where
\begin{align}\label{h1-1}
 h_1 = \sum_{e_i\in \Gamma^\E_0}|c_{0i}|^2 H^{I\E}_{ii}.
% \simeq |c_{0\alpha}|^2h_0,
\end{align}
 Now, we compute the first-order term of $\ov{\rho}_{\alpha\beta}^S$.
 For this purpose, let us rewrite Eq.(\ref{rho-S-0}) as follows,
\begin{align}\label{rho-S-0-ab}
	\ov{\rho}_{\alpha\beta}^S = \frac{\lambda }{e_\beta^S - e_\alpha^S} \sum_{\gamma=1}^{m}
\Big[H_{\alpha \gamma}^{IS} \ov{ F}_{\beta \gamma}-H^{IS}_{\gamma\beta}\ov{ F}_{\gamma\alpha} \Big].
\end{align}	
 Substituting the above-obtained zeroth-order terms $\overline{F}_{\alpha\beta}^{\rm 0th}$ into
 the rhs of Eq.\eqref{rho-S-0-ab}, one gets the following expression of $\ov{\rho}_{\alpha\beta}^S $
 up to the first-order term:
\begin{align}\label{rho-S-ab-1st-pert}
	\ov{\rho}_{\alpha\beta}^S
 \simeq \frac{\lambda H_{\alpha \beta}^{IS} h_1}{e_\beta^S - e_\alpha^S} \left(|c_{0\beta}|^2-|c_{0\alpha}|^2 \right).
\end{align}

 Finally, we compare two results obtained above, Eq.(\ref{rho-S-ab-1st-pert}) and Eq.(\ref{eq:tls-rhos}),
 the latter of which is a TLS case of the main result in Eq.(\ref{relation-explicit}).
 The two results were gotten under different conditions:
 Eq.(\ref{rho-S-ab-1st-pert}) was derived \emph{merely} under the condition of very weak coupling,
 while, Eq.(\ref{relation-explicit}) was derived under a condition that includes three requirements
 : ETH ansatz in Eq.(\ref{ETH}), Eq.(\ref{cond-h}), and largeness of $N$.
 We would remark that the above two conditions are sufficient conditions for the corresponding results, but not necessary
 conditions.
 For example,  it is possible for Eq.(\ref{eq:tls-rhos}) to hold in some cases,
 even when Eqs.(\ref{ETH}) and (\ref{cond-h}) are not fulfilled.
 In addition, none of the two conditions includes the other.

 To show consistency of the above two results, let us consider a
 case in which both conditions are satisfied.
 In fact, under Eqs.(\ref{ETH}) and (\ref{cond-h}),
 it is easy to see that $h_1$ in Eq.(\ref{h1-1})  satisfies that $h_1 \simeq h_0$.
 Then, in the weak coupling limit with Eq.(\ref{rho-aa=c0a}), Eq.(\ref{rho-S-ab-1st-pert}) is  written as
\begin{align}\label{rho-ew}
	\overline{\rho}^S_{\alpha\beta} \simeq \lambda {\eta_rh_0}
 (|c_{0\beta}|^2-|c_{0\alpha}|^2).
\end{align}
 Clearly, Eq.(\ref{rho-ew}) gives the same prediction as Eq.(\ref{eq:tls-rhos})
 in this case.

\subsection{A generic interaction}\label{sect-g-int}

 In this section,  we give a brief discussion for a generic local interaction Hamiltonian $H^I$,
 which is written as a sum of direct-product terms.
 Suppose that there are $M_{\rm LIT}$ such terms, with the subscript ``LIT'' standing for ``local interaction terms''.
 Then, $H^I$ is written as
\begin{equation}\label{HI-g}
H^I = \sum_{\nu=1}^{M_{\rm LIT}} \lambda_\nu H^{IS,\nu}\otimes H^{I\E,\nu},
\end{equation}
 where $\lambda_\nu $ are parameters and $H^{I\E,\nu}$ are local operators of the environment.
 The operators $H^{I\E,\nu}$ are assumed to satisfy the ETH ansatz, with functions $h^\nu(e)$, respectively.
 For such a generic $H^I$, the operator $F(t)$ in Eq.(\ref{F(t)}) is written as
 \begin{align}
	 F(t)=\sum_{\nu=1}^{M_{\rm LIT}} F_{\alpha\beta}^\nu(t)|\alpha\rangle\langle\beta|,
 \end{align}
where
\begin{equation}
	F_{\alpha\beta}^\nu(t) = \langle \E_\alpha(t)|H^{I\E,\nu}|\E_\beta(t)\rangle.
\end{equation}

 Following arguments similar to those given in Sec.\ref{sect-main-results},
 with appropriate generalizations,
 one may study the long-time average of this generic operator $F(t)$ and get similar results.
 More exactly, the main generalization is that Eq.(\ref{cond-h}) is now written as
 \begin{align}\label{cond-h-LIT}
	\left| \frac{1}{h_0^\nu} \Delta h^\nu \right|  \le  \frac{\epsilon_h}{M_{\rm LIT}}
\quad \text{with $\epsilon_h \ll 1$} \quad (\forall \nu),
\end{align}
 where $h_0^\nu = h^\nu(e_0)$ and
\begin{align}\label{Delta-h-LIT}
  \Delta h^\nu = \max_{e_i\in{\Gamma}^\E}|{h^\nu(e_i)- h_0^\nu}|.
\end{align}
 The final result is that, at a sufficiently large $N$,
\begin{gather}\label{tls-rho-many}
[\widetilde{H}^S,\overline{\rho}^S]\simeq 0,
\end{gather}
 where
\begin{gather} \label{HIE-ab-nu}
\widetilde{H}^S = H^S + \sum_{\nu=1}^{M_{\rm LIT}} \lambda_\nu h_{0}^\nu H^{IS,\nu}.
\end{gather}

\section{Numerical tests}\label{sect-numerical}

 In this section, we present numerical simulations that have been performed
 for checking analytical predictions given above.
 Specifically, we discuss the employed model and analytical predictions in Sec.\ref{sect-model},
% describe analytical predictions for this model in Sec.\ref{sect-analyical-prediction-model},
 and  discuss numerical simulations in Sec.\ref{sect-numerical-simul}.

\subsection{The model}\label{sect-model}

 In numerical simulations, we employ a TLS as the central system $S$
 and one defect Ising chain as the environment $\E$.
 The TLS has a self-Hamiltonian written as
\begin{gather}\label{HS-model}
 H^S = q_s S^z,
\end{gather}
 where $q_s$ is a parameter and $S^z$ indicates the $z$-component Pauli matrix divided by $2$.

 The defect Ising chain is composed of a number $N$ of $\frac{1}{2}$-spins
 lying in an inhomogeneous transverse field, whose Hamiltonian is written as
\begin{equation}
H^\E = B_x\sum_{l=1}^NS_l^x + d_1 S_1^z + d_5 S_5^z  +  J_z\sum_{l=1}^NS_l^z S^z_{l+1},
\end{equation}
 where $S^x_l$ and $S^z_l$ indicate Pauli matrices divided by $2$ at the $l$-th site.
 Here, $B_x$, $J_z$, $d_1$, and $d_5$ are parameters, which are adjusted
 such that the defect Ising chain is a quantum chaotic system.
 That is, for levels not close to edges of the energy spectrum,
 the nearest-level-spacing distribution $P(s)$
 is close to the Wigner-Dyson distribution $P_W(s) = \frac{\pi}{2}s\exp(-\frac{\pi}{4}s^2)$,
 the latter of which is almost identical to the prediction of RMT
 \cite{haake2013quantum,casati1980connection,bohigas1984characterization}.
 Exact values of the parameters used are $B_x = 0.9, J_z = 1.0$,  $d_1 = 1.11$, and $ d_5 = 0.6$;
 and $N$ is between $10$ and $13$.
 In our numerical  computation of EFs, the periodic boundary condition was implied
 and the so-called Krylov-space method was used.

 The TLS is coupled to the $k$-th spin of the defect Ising chain.
 We have studied two specific forms of the local interaction Hamiltonian, indicated as $H^I_{(1)}$ and $H^I_{(2)}$,
\begin{subequations}\label{Hint-model}
\begin{align}
&	H^I_{(1)} = \lambda S^x\otimes S_k^x,
\\ &  H^I_{(2)} = \lambda (S^x + S^z)\otimes S_k^x.
\end{align}
\end{subequations}
 Their difference  lies in that the TLS part of $H^I_{(1)}$ has no overlap with $H^S$ in Eq.(\ref{HS-model}),
 while, $H^I_{(2)}$ has some.
  According to Eqs.(\ref{Delta-S}) and \eqref{eta-dr},
  one finds that $\Delta_S=q_s$, $\eta_d= 0$ and $\eta_r=\frac{1}{2q_s}$ for $H^I_{(1)}$,
 and $\eta_d=-\frac{1}{q_s}$
 and $ \eta_r=\frac{1}{2q_s}$ for $H^I_{(2)}$.
 Numerically, we have checked that the ETH ansatz is applicable to local operators in the defect Ising chain
 (see Appendix \ref{sect-numerical-eth}).

%\subsection{Analytical predictions}\label{sect-analyical-prediction-model}

 \begin{figure}[ht]
	\includegraphics[width=1.0\linewidth]{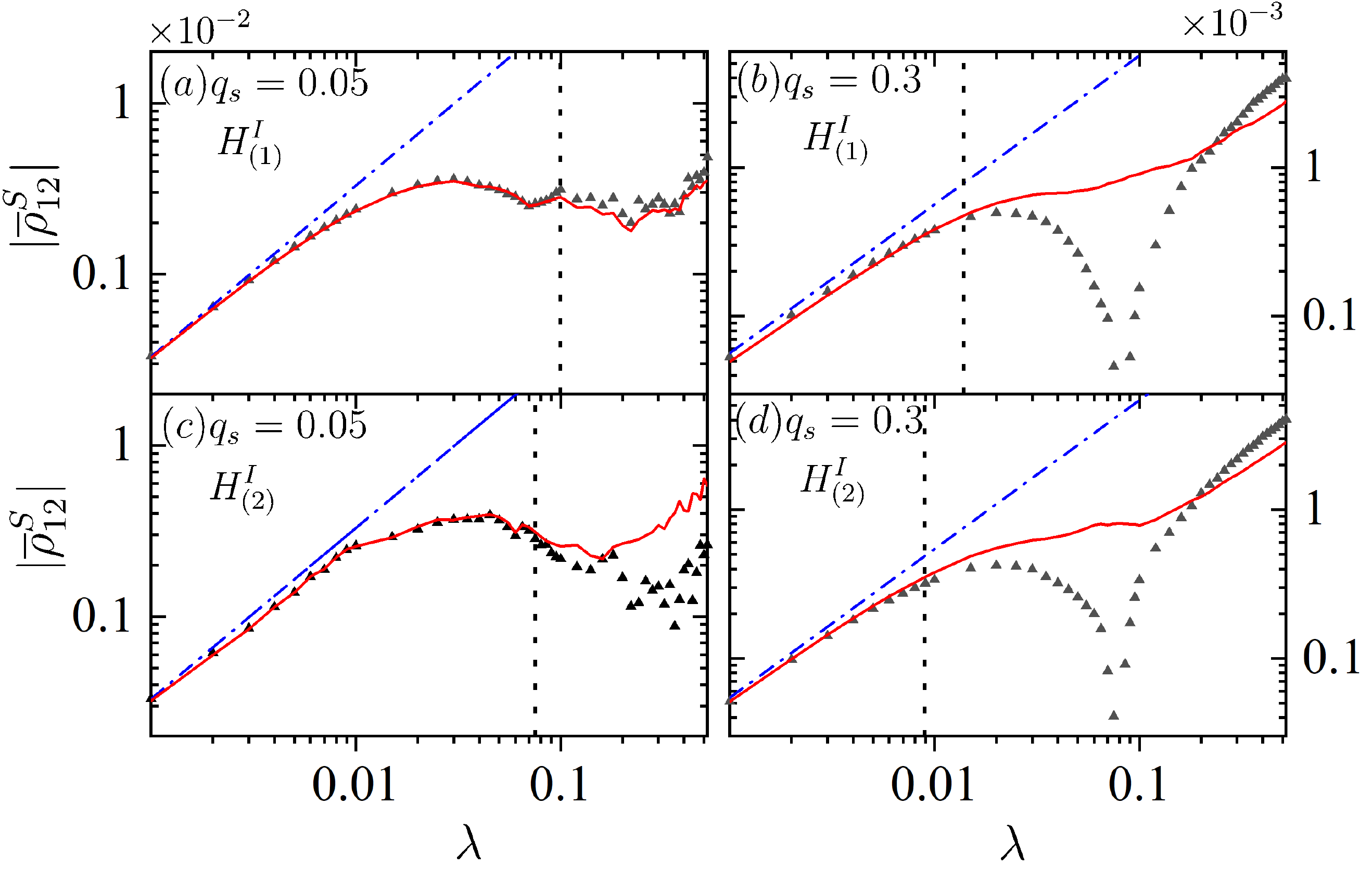}
	\caption{Values of $|\ov{\rho}^S_{12}|$ (triangles) vs the coupling strength $\lambda$
 in the logarithm scale, under two interaction Hamiltonians of $H^I_{(1)}$ and $H^I_{(2)}$.
 Left panels: $q_s=0.05$, and right panels: $q_s=0.3$.
 The solid lines (red) represent predications of the main result Eq.\eqref{eq:tls-rhos},
 and the dashed-dotted lines (blue) show predications of Eq.(\ref{rho-S-ab-1st-pert})
 for very weak couplings.
 The vertical  dot lines (black) indicate positions of $\lambda_c$,
 which were computed by Eq.\eqref{Dh-h0-eR} with $\epsilon_c =0.1$.
Parameters: $(c_{01},c_{02}) = (0.51,0.86)$,  $N=13$, $e_0=-1.2$, $\delta e_0 = 0.1$, and $k=7$.
 }
	\label{fig:rhos_interm}
\end{figure}

 Below, we discuss predictions for properties of the long-time-averaged RDM element $\ov \rho^S_{12}$ of the TLS,
 which are given by analytical results of previous sections.
 We discuss in the increasing order of the interaction strength $\lambda$.

 \vspace{0.2cm}
 (1) Regime of very small $\lambda$ (weak coupling limit).
 \\ As discussed in Sec.\ref{sect-weak-coupling},
 $\ov \rho^S_{12}$ should satisfy Eq.(\ref{rho-S-ab-1st-pert}) at very small $\lambda$.
 Since the ETH ansatz is applicable to the defect Ising chain, when Eq.(\ref{cond-h}) is satisfied,
 this prediction coincides with Eq.(\ref{eq:tls-rhos}),
 which is the TLS case of the main result in Eq.\eqref{relation-explicit}.

 \vspace{0.2cm}
 (2) Regime of small but not very small $\lambda$.
\\ (a) Eq.(\ref{cond-h}) being valid at $\lambda =0$.
 \\ In this case, Eq.(\ref{cond-h}) is also valid at small $\lambda$.
 As a result,  $\ov \rho^S_{12}$ should satisfy the main prediction Eq.(\ref{eq:tls-rhos}) at sufficiently large $N$.
\\ (b) Eq.(\ref{cond-h}) being invalid at $\lambda =0$.
 \\ In this case, there is no definite analytical prediction for $\ov \rho^S_{12}$ beyond the weak coupling limit.

 \vspace{0.2cm}
 (3) Regime of $\lambda$ below $\lambda_h$ with Eq.(\ref{cond-h}) valid.
 \\  As discussed in Sec.\ref{sect-relevance-IS}, Eq.(\ref{eq:tls-rhos}) is applicable for $\lambda$ below $\lambda_h$.
 The value of $\lambda_h$, which satisfies Eq.(\ref{rela-lamb-h}), is expected to increase with increasing $N$
 if $\mu <1$,
 while, increase with decreasing $\Delta_S$.

 As discussed previously, Eq.(\ref{cond-h}) belongs to a sufficient, but not necessary,
 condition for validity of  Eq.(\ref{eq:tls-rhos}).
 This implies that Eq.(\ref{eq:tls-rhos}) might be useful even beyond $\lambda_h$.
 To directly study validity of Eq.(\ref{eq:tls-rhos}), one may compute the value of $\lambda$, indicated by $\lambda_c$,
 at which the relative error first reaches some small parameter indicated by $\epsilon_c$
 when $\lambda$ increases from $0$,
\begin{align}\label{Dh-h0-eR}
 \left| \frac{\ov \rho^S_{12} - \ov \rho^S_{12,{\rm th}}}{\ov \rho^S_{12}} \right|_{\lambda = \lambda_c} = \epsilon_c,
\end{align}
 where $\ov \rho^S_{12}$ indicates the exact value of the RDM element
 and $\ov \rho^S_{12,{\rm th}}$ is for the prediction of Eq.(\ref{eq:tls-rhos}).

 We have no definite analytical prediction for behaviors of $\lambda_c$.
 It seems reasonable to expect that, at least in some cases,
 $\lambda_c$ may show some behavior qualitatively similar to that of $\lambda_h$ as discussed
 above in predication (3).

\begin{figure}[ht]
	\includegraphics[width=1.0\linewidth]{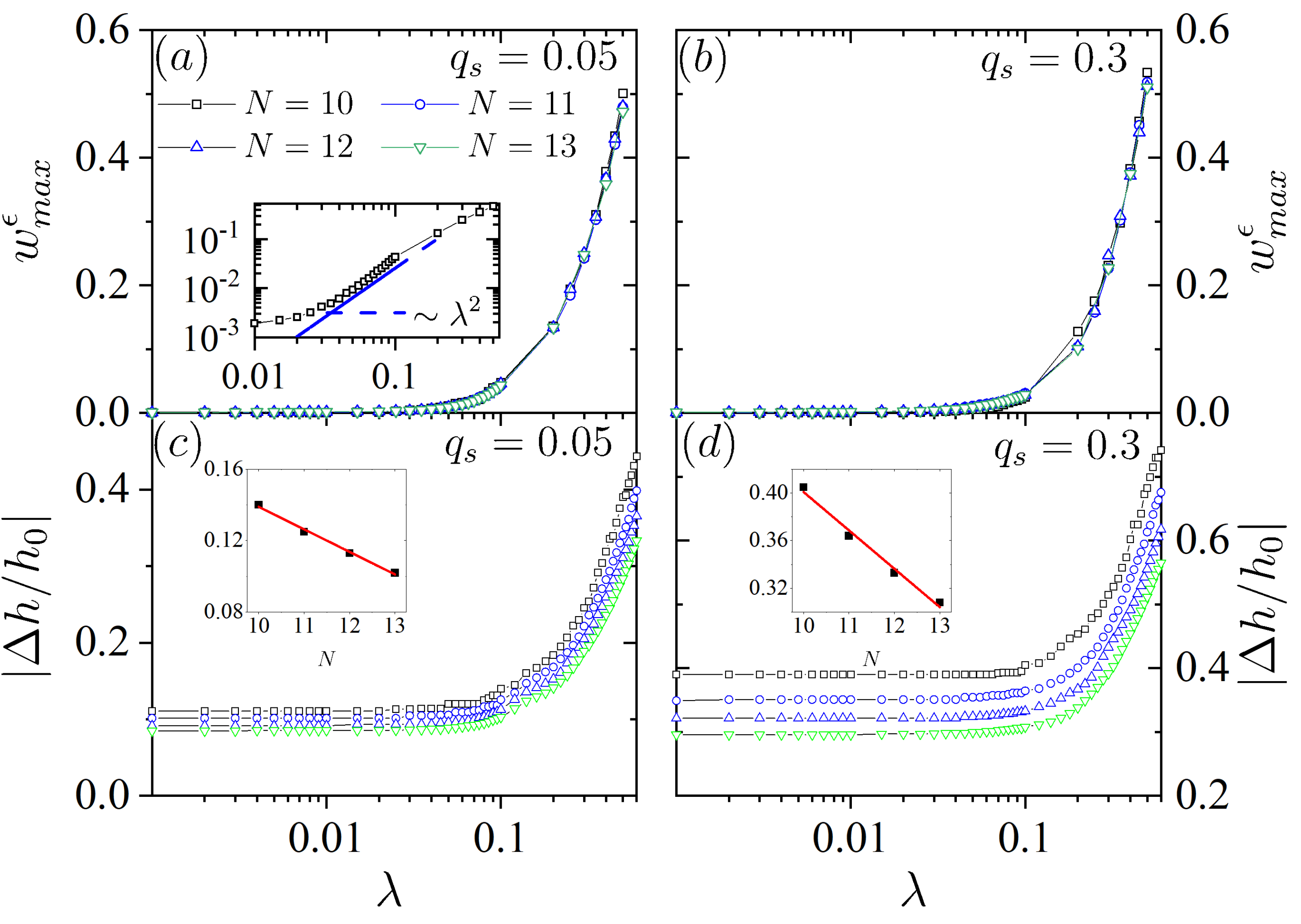}
	\caption{ (a) Variation of $w_{\rm max}^\epsilon$ vs  $\lambda$, with $\lambda$ in the logarithm scale,
  for $q_s=0.05$, $N\in [10,13]$, and $\epsilon =0.05$.
 The interaction Hamiltonian is $H_{(1)}^I$.
 Inset: both axes in the logarithm scale, showing that $w_{\rm max}^\epsilon$ has approximately a
 $\lambda^2$-behavior in the middle regime of $\lambda$.
  (b) Similar to (a), but for $q_s=0.3$.
  (c) Variation of  $|\Delta h/h_0|$ vs $\lambda$ for $q_s =0.05$.
  Inset: $|\Delta h/h_0|$ vs $N$  at $\lambda =0.1$, showing a $1/N$ behavior, as predicted by
  Eq.\eqref{eq:fluc-h-estimate}.
  (d) Similar to (c), but for $q_s =0.3$, with some difference in the scale of the vertical axis.
 Other parameters are the same as in Fig.\ref{fig:rhos_interm}, except that $e_0$ is determined by
 keeping $e_0/N$ constant.
 }
 \label{fig:eth-ratio}
\end{figure}

\subsection{Numerical simulations}\label{sect-numerical-simul}

 We have numerically checked the above predictions
 for various values of the parameters concerned.
 The environmental initial state was taken as a typical state within an energy shell $\Gamma^{\cal E}_0$,
 which is given by $e_0=-1.2$ and $\delta e_0 = 0.1$.

 Two values of $q_s$ has been studied, namely, $q_s =0.3$ and $0.05$.
 For $q_s = 0.3$,  we found that $|\Delta h/ h_0| \simeq 0.6$ at $\lambda =0$ and  $N=13$,
 implying invalidity of Eq.(\ref{cond-h}).
 With $q_s$ changed to $q_s = 0.05$, we found $|\Delta h/ h_0| \simeq 0.1$,
 implying validity of Eq.(\ref{cond-h}).
 In both cases,  $h_1 \simeq h_0$.

 Variations of $|\ov \rho^S_{12}|$ versus the interaction strength $\lambda$ are shown in
 Fig.\ref{fig:rhos_interm}, for the above-mentioned two values of $q_s$ and
 for the two interaction Hamiltonians in Eq.(\ref{Hint-model}).
 One sees that there is in fact no qualitative difference between results for the two interaction Hamiltonians.
 In the computation of the rhs of Eq.(\ref{eq:tls-rhos}),
 exact values of $\ov{\rho}^S_{11}$ and $\ov{\rho}^S_{22}$ were used.
 In agreement with prediction,
 both the main result of Eq.(\ref{eq:tls-rhos}) (solid lines) and
 the weak-coupling prediction of Eq.(\ref{rho-S-ab-1st-pert})  (dashed-dotted lines)
 work well  at very small $\lambda$, more exactly, at $\lambda$ around $0.001$ and smaller.
 Consistently, the mean nearest-level spacing
 of the total system was found about $7.3\times 10^{-4}$ in the considered energy region at $N=13$.

 With $\lambda$ increased above $0.001$, as expected, the weak-coupling predictions (dashed-dotted lines in blue)
 gradually deviate from the exact values of $|\ov{\rho}^S_{12}|$ (triangles).
 Meanwhile, consistent with the prediction of (2)(a),
 for $q_s = 0.05$ with Eq.(\ref{cond-h}) valid,
 predictions of Eq.(\ref{eq:tls-rhos}) (solid lines in red) remain close to the triangles,
 up to $\lambda \sim 0.1$.
 It is of interested to note that, even in the case of $q_s = 0.3$ with Eq.(\ref{cond-h}) unsatisfied,
 predictions of Eq.(\ref{eq:tls-rhos}) remains valid up to $\lambda \simeq 0.01$.

 To get further understanding for the above-discussed behaviors of $|\ov \rho^S_{12}|$,
 we have studied variation of the maximum width $w^\epsilon_{max}$,
 which is responsible to the $\lambda$-dependence of
 the width $\delta e (= \delta e_0+2q_s+w_{max}^\epsilon)$ of $\Gamma^\E$,  versus $\lambda$,
 as well as variation of  $|\Delta h/ h_0|$ (Fig.\ref{fig:eth-ratio}).
 It is seen that, at $q_s =0.05$ and $N=13$, the value of $w^\epsilon_{\rm max}$ keeps small
 for small $\lambda$
 and begin to increase fast around $\lambda =0.1$;
 and, consistently, $|\Delta h/ h_0|$ (triangles down) behaves in a similar way.
 Similar behaviors are seen at $q_s =0.3$  and $N=13$,
 except that $|\Delta h/ h_0|$ is already large at $\lambda =0$.

\begin{figure}[ht]
	\includegraphics[width=1.0\linewidth]{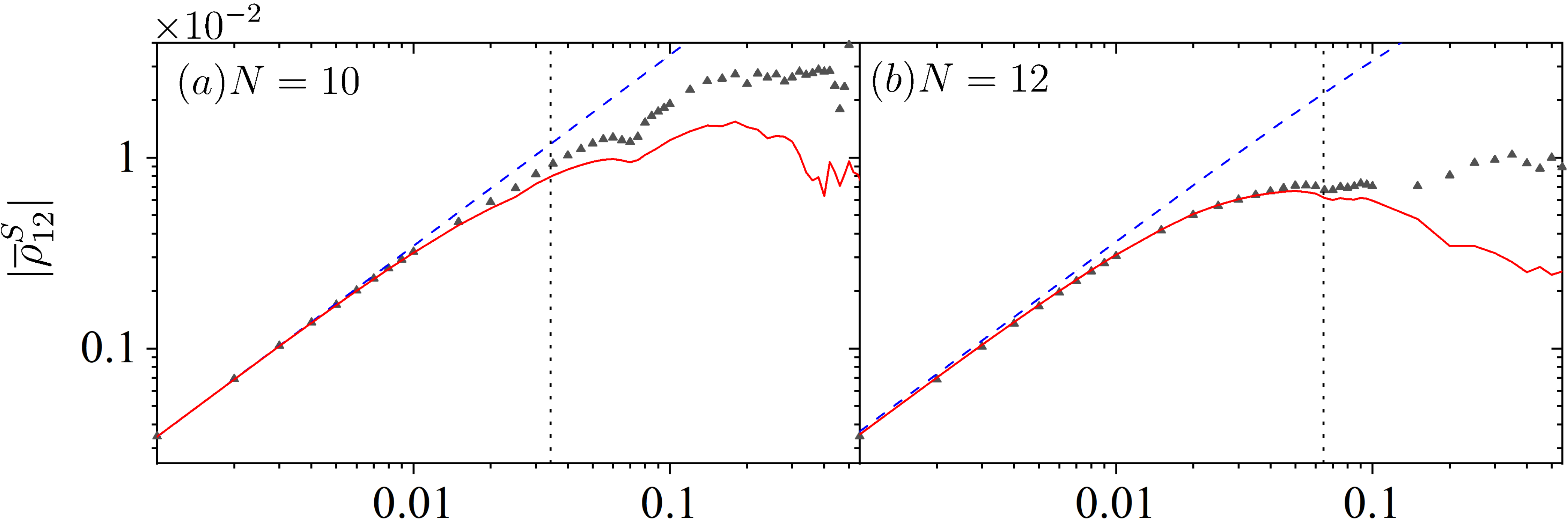}
	\caption{Similar to  Fig.\ref{fig:rhos_interm} (a), but, for $N=10$ and $12$.
}
	\label{fig:rhos-size}
\end{figure}

 We have also studied impact of the particle number $N$.
 As seen in Fig.\ref{fig:eth-ratio}, the width $w^\epsilon_{\rm max}$ is almost independent of $N$
 for $N$ from $10$ to $13$, which implies a negligible value of $\mu$, i.e., $\mu \approx 0$.
 Meanwhile, the value of $|\Delta h/h_0|$ decreases with increasing $N$,
 in agreement with a prediction of Eq.\eqref{eq:fluc-h-estimate} that
 $\Delta h$ may scale as $1/N$ at a fixed value of $e_0/N$
 [insets of Fig.\ref{fig:eth-ratio}(c) and (d)].
 Moreover, with $\mu \approx 0$,
 according to prediction (3), $\lambda_h$ may increase with increasing $N$;
 in other words, Eq.\eqref{eq:tls-rhos} may work better at larger $N$,
 which is seen by comparing Fig.\ref{fig:rhos-size} and Fig.\ref{fig:rhos_interm}.

 Finally, we discuss numerically obtained values of $\lambda_h$ for validity of Eq.(\ref{cond-h})
 and values of $\lambda_c$ for practical use of Eq.\eqref{eq:tls-rhos}.
 The values of $\lambda_h$  may be directly gotten from Fig.\ref{fig:eth-ratio}.
 Taking $\epsilon_h =0.1$, we found that
 Eq.(\ref{cond-h}) is valid for no value of $\lambda$ at $q_s =0.3$ and $N=10,11,12,13$,
 which is indicated as $\lambda_h={\rm none}$ in  Table \ref{table-lambda};
 and, similarly, for $q_s=0.05$ and $N=10$.
 Thus, $\lambda_h$ has definite values only at $q_s=0.05$ and $N=11,12,13$.
 If the restriction of $\epsilon_h =0.1$ is loosed a little, i.e., if taking $\epsilon_h$ larger than $0.1$ but still small
 (e.g., $0.15$),
 $\lambda_h$ may have definite values in more cases, which is clear from Fig.\ref{fig:eth-ratio}(c).
 In all the cases in which $\lambda_h$ has definite values for Eq.(\ref{cond-h}),
 we found that larger value of $\lambda_h$ corresponds to larger value of $N$, meanwhile,
 larger value of $\lambda_h$ corresponds to smaller value of $q_s$,
 in agreement with  prediction (3).

 Values of $\lambda_c$ were computed by making use of Eq.(\ref{Dh-h0-eR}) with $\epsilon_c =0.1$.
 At $q_s=0.05$  with Eq.(\ref{cond-h}) valid, as seen in Table \ref{table-lambda},
 $\lambda_c$ increases with increasing $N$ and is close to $\lambda_h$ for $N=11,12,13$.
 But, at $q_s=0.3$ with Eq.(\ref{cond-h}) invalid, $\lambda_c$ shows a quite complicated behavior;
 more exactly, it does not increase monotonically with $N$ and is unexpectedly large at $N=10$.

%  \begin{figure}[ht]
% 	\includegraphics[width=1.0\linewidth]{table.pdf}
% 	\caption{figure}
% \label{figure}
%  \end{figure}
\begin{table}
	\centering
\begin{tabular}{c  c  c  c  c }
	\hline
	 &$N=10$ & $N=11$ & $N=12$ & $N=13$ \\ [1.5ex]
	 \hline
	 $\lambda_h(q_s=0.05)$ &  None & 0.04 &0.07 & 	 0.1\\
	 $\lambda_h(q_s=0.3)$ &  None & None & None & 	 None \\
	 $\lambda_c (q_s=0.05)$ &  0.025 & 0.04 & 0.07 & 0.1\\
	 $\lambda_c (q_s=0.3)$ &0.1   & 0.015 & 0.025 & 0.015\\
	 \hline
\end{tabular}
\caption{Values of $\lambda_c$ and $\lambda_h$, obtained with $\epsilon_h = \epsilon_c =0.1$, the interaction Hamiltonian is $H_{(1)}^I$.}
\label{table-lambda}
\end{table}

\section{Conclusions and discussions}\label{sect-conclusion}

 In this paper, the long-time averaged RDM has been studied for a generic small central system $S$ with $m$ levels,
 which is locally coupled to a large many-body chaotic environment $\E$,
 with the total system undergoing a Schr\"{o}dinger evolution.
 Beside largeness of the particle number of $\E$, the only restriction is that
 the environmental part of the interaction Hamiltonian
 satisfies the ETH ansatz, with the diagonal term in the ansatz [namely the function $h(e)$] approximately a constant
 within the energy region of relevance.
 For such a total system, on the eigenbasis of the central system,
 $\frac 12 (m-1)(m+2)$ approximate relations have be derived among elements of its steady states (if existing).

 The above-discussed relations imply that the steady RDM should be commutable
 with a renormalized Hamiltonian $\ww{H}^S$ of the central system,
 which includes certain averaged impact of the system-environment interaction.
 As a consequence, decoherence happens on the eigenbasis of the renormalized Hamiltonian,
 even under a system-environment interaction that is dissipative for the original Hamiltonian $H^S$,
 and leads to a PBS given by  the eigenbasis of $\ww{H}^S$.
 This enriches analytical knowledge about PBS for systems under nonweak and dissipative
 system-environment interactions, which had been previously observed numerically in some specific models
 (see, e.g., Ref.\cite{wang2012preferred}).
 \footnote{We would note a difference between the type of models studied in this paper and
 the spin-boson models used in
 Refs.\cite{lee2012noncanonical,addis2014coherence,roszak2015decoherence,zhang2015role, guarnieri2018steady}.
 That is, in a spin-boson model, the environment (the bosons) is not a quantum
 chaotic system and the ETH ansatz is usually inapplicable.
 Due to this difference, even if a PBS may exist in a spin-boson model,
 the mechanism should be quite different from that discussed in this paper. }
 Moreover, results of this paper give an explicit way of constructing renormalized Hamiltonian for PBS.

 In fact, renormalized Hamiltonian is also used in a standard master-equation approach to RDM.
 There, at an initial stage before the derivation begins,
 the self-Hamiltonian of the central system is taken as certain renormalized Hamiltonian,
 which we indicate as $H^S_{\rm mas}$ with ``mas'' standing for ``master equation'', given by
 $H^S_{\rm mas}=H^S+\lambda H^{IS}\text{tr}(H^{I\E}\rho^\E_{\rm th})$, where $\rho^\E_{\rm th}$
 denotes a thermal state of the environment.
 Under the ETH ansatz and the condition in Eq.(\ref{cond-h}),
 $H^S_{\rm mas}$ has almost the same expression as $\ww H^S$  in Eq.(\ref{Hs-normal}),
 if the state $\rho^\E_{\rm th}$ lies effectively within the energy shell $\Gamma^\E$.

 However, there is a big difference between the physical meanings of $H^S_{\rm mas}$ and $\ww H^S$.
 In fact, in our approach, the operator $\ww H^S$ is derived by faithfully
 taking the long-time average over the overall Schr\"{o}dinger evolution;
 and it indicates the existence of a PBS, if the RDM may approach a steady state.
 While, in the master-equation approach,
 the operator $H^S_{\rm mas}$ is mainly employed for the sake of convenience in derivation,
 though with deep physical intuition lying behind it.
 Only after a certain type of analytical solution to a derived master equation is found,
 which is usually a hard task except in some special models,
 could it become clear whether $H^S_{\rm mas}$ may indeed be of relevance to a PBS.
 Moreover, as an approach based a perturbative treatment,
 validity of the master-equation approach at long times is a subtle issue.

 Finally, we would mention that,
 beside the field of decoherence, results of this paper may also be useful in other fields
 in which properties of steady states of small and open quantum systems are of relevance,
 such as quantum thermodynamics
 \cite{binder2019thermodynamics, goold2016role,wang2012statistical,wang2018decoherence}.

%  Finally, we give a brief discussion on the condition Eq.(\ref{cond-h}), under which our main results are derived.
%  \\ (i) As long as $w^\epsilon_{\rm max}$ is notably smaller than $( \delta e_0/2+\Delta_S)$,
%  validity of Eq.(\ref{cond-h}) is independent of the interaction strength.
%  This implies that our main results may be valid much above the weak coupling limit,
%  even to some nonweak interaction regime, as illustrated by our numerical simulations.
%  \\ (ii) According to numerical simulations with some analytical analysis
%  \cite{kim2014testing,brenes2020eigenstate, d2016quantum,brenes2020low},
%  $h(e)$ is usually a function of  per-site energy, i.e.
%  \begin{align}\label{he-norma}
% 	 h(e_i) =\widetilde{h}(e_i/N),
%  \end{align}
%  where $\widetilde{h}(x)$ is some smooth function of $x$.
%  This expression gives that
%  \begin{align}
% 	\label{eq:eth-diag-diff}
% 	 |h(e_i)-h(e_j)| = \left| \ww h'(e_i/N) \right| \frac{|e_i-e_j|}{N} + O(1/N^2),
%  \end{align}
%  and, as a result, $\Delta h \sim 1/N$.
%  This implies that one may make the condition (\ref{cond-h}) fulfilled by increasing $N$,
%  if, for example, the width $w^\epsilon_{\rm max}$ may be kept not large, meanwhile,
%  parameters are arranged such that $\Delta_S, (e_0/N)$, and $ \delta e_0$ are almost
%  independent of $N$.

\acknowledgments

% The authors are grateful to J. Gong and G. Benenti for valuable suggestions.
 This work was partially supported by the Natural Science Foundation of China under Grant
 No.~11535011, 11775210, and 12175222. JW are supported by the Deutsche Forschungsgemeinschaft (DFG) within the Research Unit FOR 2692
 under Grant No. 397107022 (GE 1657/3-2).

\appendix
\renewcommand{\thefigure}{\thesection. \arabic{figure}}
\setcounter{figure}{0}
\section{Derivation of Eq.\eqref{drhot}}
\label{append1}
 In this appendix, we derive Eq.\eqref{drhot}.
 Using Eq.\eqref{rho-abt}, the time evolution of the RDM is written as
\begin{equation}
i\hbar \frac{d\rho_{\alpha\beta}^S(t)}{dt} = i\hbar\dt \langle \E_\beta(t)|\E_\alpha(t)\rangle  = A_1 +A_2,
\end{equation}
where
\begin{subequations}
\begin{align}
	A_1 &= i\hbar \left( \dt \langle \E_\beta(t)| \right) |\E_\alpha(t)\rangle, \\
	A_2 & = i\hbar\langle \E_\beta(t)| \left( \dt |\E_\alpha(t)\rangle \right).
\end{align}
\end{subequations}
Making use of Eq.\eqref{alpha-t}, one finds that
\begin{align}
	\label{f1-1}
	A_1 = -\sum_{\gamma}\langle
\E_\gamma(t)|H_{\gamma \beta }|\E_\alpha(t)\rangle.
\end{align}
 From Eqs.\eqref{HI} and \eqref{H-subspace}, one gets that
\begin{align}
 H_{\alpha\beta} = \langle\alpha|H|\beta\rangle=e^S_\alpha\delta_{\alpha\beta}
 +\lambda H^{IS}_{\alpha\beta} H^{I\E} + H^\E\delta_{\alpha\beta}.
\end{align}
 Then, we write Eq.\eqref{f1-1} as,
\begin{align}
	A_1 =& -e_\beta^S\langle \E_\beta(t)|\E_\alpha(t)\rangle -\lambda \sum_{\gamma}
 H^{IS}_{\gamma\beta}\langle
\E_\gamma(t)|H^{I\E}|\E_\alpha(t)\rangle\nonumber \\& -\langle \E_\beta(t)|H^\E|\E_\alpha(t)\rangle.
\end{align}
 Noting Eqs.\eqref{rho-abt} and \eqref{HIE-ab}, the above equality gives that \begin{align}
	A_1 = -e_\beta^S \rho^S_{\alpha\beta}(t) -\lambda \sum_\gamma H^{IS}_{\gamma\beta}F_{\gamma\alpha}(t)
	-\langle \E_\beta(t)|H^\E|\E_\alpha(t)\rangle.
\end{align}
 Similarly, one finds
\begin{equation}
	A_2 = e_\alpha^S\rho_{\alpha\beta}^S(t) + \lambda \sum_\gamma H^{IS}_{\alpha\gamma}F_{\beta\gamma}(t)
+\langle \E_\beta(t)|H^\E|\E_\alpha(t)\rangle.
\end{equation}
 Putting the above results together, one gets Eq.\eqref{drhot}.

% \section{Some Properties of the environmental \emph{branches}}
% \label{appen-env}
% From Eq.\eqref{fai}, we write the full expression of the expanding coefficients of the environmental \emph{branches}
% \begin{align}
% 	f_{\alpha i}(t)=\langle i|\E_\alpha(t)\rangle =\frac{1}{\N_0}\sum_{k\in\Gamma_0^\E,n,\alpha^\prime}c_{0\alpha^\prime}c_{0k}e^{-iE_nt}C_{\alpha^\prime k}^{n*}C_{\alpha i}^n,
% \end{align}
% by introducing the property of typical state $\langle c_{0i}c_{0j}\rangle=\delta_{ij}$, one encounters
% \begin{align}
% 	f_{\beta i}^*(t)&f_{\alpha j}(t)\simeq \frac{1}{N_{\Gamma_0^\E}}\sum_{k,l\in\Gamma_0^\E,nm,\alpha^\prime\beta^\prime}c_{0\alpha^\prime}c_{0\beta^\prime}^*c_{0k}c_{0l}^*\times\nonumber\\
% 	&e^{-i(E_n-E_m)t}C_{\alpha j}^nC_{\beta i}^{m*}C_{\alpha^\prime k}^{n*}C_{\beta^\prime l}^{m},\nonumber\\
% 	\qquad &\simeq \frac{1}{N_{\Gamma_0^\E}}\sum_{k\in\Gamma_0^\E,nm,\alpha^\prime\beta^\prime}c_{0\alpha^\prime}c_{0\beta^\prime}^*e^{-i(E_n-E_m)t}\times \nonumber
% 	\\ &\qquad C_{\alpha j}^nC_{\beta i}^{m*}C_{\alpha^\prime k}^{n*}C_{\beta^\prime k}^{m},
% \end{align}
% while the non-degeneracy condition leads to
% \begin{align}
% 	\overline{f_{\beta i}^*f_{\alpha j}}:=	\overline{f_{\beta i}^*(t)f_{\alpha j}(t)}&\simeq \frac{1}{N_{\Gamma_0^\E}}\sum_{k\in\Gamma_0^\E,n,\alpha^\prime\beta^\prime}c_{0\alpha^\prime}c_{0\beta^\prime}^*\times \nonumber\\
% 	&C_{\alpha j}^nC_{\beta i}^{n*}C_{\alpha^\prime k}^{n*}C_{\beta^\prime k}^{n}.
% \end{align}

\section{Scaling of the fluctuation operator}
\label{append3:norm-bound-eth}

 In this appendix, we show that the main $N$-scaling behavior of the fluctuation operator $\Delta$,
 \begin{equation}
\Delta = \sum_{\alpha \beta} \Delta_{\alpha \beta} |\alpha\ra \la \beta|,
 \end{equation}
 is an exponential decay with increasing $N$.
 For this purpose, let us compute the Frobenius norm of $\Delta$,
\begin{align}
||\Delta||_F^2 &=\sum_{\alpha\beta}|\Delta_{\alpha\beta}|^2.
\end{align}
 Making use of Eq.(\ref{Delta-ab}), direct derivation shows that
\begin{align}
 ||\Delta||_F^2 &=\sum_{\alpha\beta} \left|\sum_{ij}g(e,\omega)e^{-S(e)/2}
\overline{f_{\beta i}^*f_{\alpha j}}R_{ij} \right|^2\nonumber \\
&=\sum_{\alpha\beta}\sum_{i j i' j'}g(e,\omega)e^{-S(e)/2}
\overline{f_{\beta i}^*f_{\alpha j}}R_{ij}\nonumber\\
&\quad\quad\times {g}^*(e',\omega')e^{-S(e')/2}
\overline{f_{\beta i'}f_{\alpha j'}^*}\ {R}^*_{i'j'}.
\end{align}
 Note that $g(e,\omega)={g}^*(e,-\omega)$ and $R_{ij}={R}^*_{ji}$.
 To proceed, let us discuss the statistical average of $||\Delta||_F^2$,
 taken over the random variables $R_{ij}$, which is indicate by
 $\langle \cdot\rangle$.
 This averaging procedure results in that   \cite{murthy2019bounds,d2016quantum}
\begin{align}
\label{eq:eth-random}
  \langle R_{ij} {R}^*_{i'j'}\rangle=\delta_{ii'}\delta_{jj'}+\delta_{ij'}\delta_{i'j},
\end{align}
  and, as a consequence,
\begin{align}
 \la ||\Delta||_F^2 \ra &= \sum_{ij}|g(e,\omega)|^2e^{-S(e)}\nonumber\\
 &\quad\quad \times \left(\sum_{\alpha\beta}\big|\overline{f_{\beta i}^*f_{\alpha j}}\big|^2
 +\overline{f_{\beta i}^*f_{\alpha j}}\cdot\overline{f_{\beta j}f_{\alpha i}^*} \right)\nonumber\\
 &\le 2\sum_{ij}|g(e,\omega)|^2e^{-S(e)}\sum_{\alpha\beta}\big|\overline{f_{\beta i}^*f_{\alpha j}}\big|^2\nonumber\\
&\le 2 \max_{ij}(|g(e,\omega)|^2e^{-S(e)})\sum_{ij}\sum_{\alpha\beta}\big|
\overline{f_{\beta i}^*f_{\alpha j}}\big|^2. \label{<DF>-1}
% &= \frac{2}{I_0}\max_{ij}(|g(e,\omega)|^2e^{-S(e)}),
\end{align}

 To compute $\overline{f_{\beta i}^*f_{\alpha j}}$, we make use of the fact that
 $f_{\alpha j} = \la \alpha j|\Psi(t)\ra$.
 This gives that
\begin{align}
	&\overline{f_{\beta i}^*f_{\alpha j}}
	=\overline{\langle \Psi(t)|\beta i\rangle\langle \alpha j|\Psi(t)\rangle}\nonumber\\
	&=\sum_{mn} \langle \Psi(0)|n\rangle \langle n|\beta i\rangle\overline{e^{i(E_n-E_m)t}}\langle
\alpha j|m\rangle \langle m|\Psi(0)\rangle.
\end{align}
 Note that the environment $\E$, as a quantum chaotic system, has a nondegenerate spectrum.
 Under a generic system-environment interaction, the spectrum of the total system is nondegenerate, too.
 Then, one has $\overline{e^{i(E_n-E_m)t}}=\delta_{mn}$ and, as a result,
\begin{align}\label{ovff-Psi}
	&\overline{f_{\beta i}^*f_{\alpha j}}
=\sum_{n} \langle \Psi(0)|n\rangle \langle n|\beta i\rangle
 \la \alpha j|n\rangle \langle n|\Psi(0)\rangle.
\end{align}
 This gives that
\begin{align}\notag
	\left| \overline{f_{\beta i}^*f_{\alpha j}} \right|^2
 & =\sum_{n m} \langle \Psi(0)|n\rangle \langle n|\beta i\rangle  \langle \beta i|m\rangle \langle m|\Psi(0)\rangle
 \\ & \times   \langle \Psi(0) | m\rangle  \la m|\alpha j \rangle \la \alpha j|n\rangle \langle n|\Psi(0)\rangle .
\end{align}

 Then, making use of the completeness of the basis of $|\alpha\ra $ and that of $|i\ra$, one gets that
\begin{align}\label{sum-f-L0}
	& \sum_{\alpha \beta i j}\left| \overline{f_{\beta i}^*f_{\alpha j}} \right|^2 = \frac{1}{L_0},
\end{align}
 where $L_0$ is the so-called participation function of the initial state $|\Psi(0)\ra$, defined by
\begin{align}
L_0 = \frac 1{\sum_{n } \left| \langle \Psi(0)|n\rangle \right|^4}.
\end{align}
 As is known, $L_0$ gives a measure to the localization length, i.e., to the number  of those levels $E_n$
 that are effectively occupied by the state $|\Psi(0)\ra$.
 For a large environment and an initial shell not extremely narrow, the value of $L_0$ is large.

 Substituting Eq.(\ref{sum-f-L0}) into Eq.(\ref{<DF>-1}),
 we get an upper bound to the averaged norm $\la ||\Delta||_F^2 \ra$, i.e.,
\begin{align}
	\label{norm-Delta-ub}
 \la ||\Delta||_F^2 \ra  &\le \frac{2}{L_0}\max_{ij}|g(e,\omega)|^2e^{-S(e)} 
 \sim  \frac{2 N^{2\gamma}}{L_0}  e^{-S(e)}.
\end{align}
 Since the averaging procedure does not change the $N$-scaling behavior of the norm $||\Delta||_F^2$
 and the exponential-decay term $e^{-S(e)/2}$ already exists in the
 exact expression of $\Delta_{\alpha \beta}$ in Eq.(\ref{Delta-ab}), from Eq.(\ref{norm-Delta-ub})
 one sees that the  $N$-scaling behavior of fluctuation operator $\Delta$ should be
 dominated by the exponential decay $e^{-S(e)/2}$.
% (Besides, the fluctuation operator may be further reduced by the initial spreading $L_0$.)

\section{Verification of ETH ansatz}\label{sect-numerical-eth}
\begin{figure}
	\includegraphics[width=1.0\linewidth]{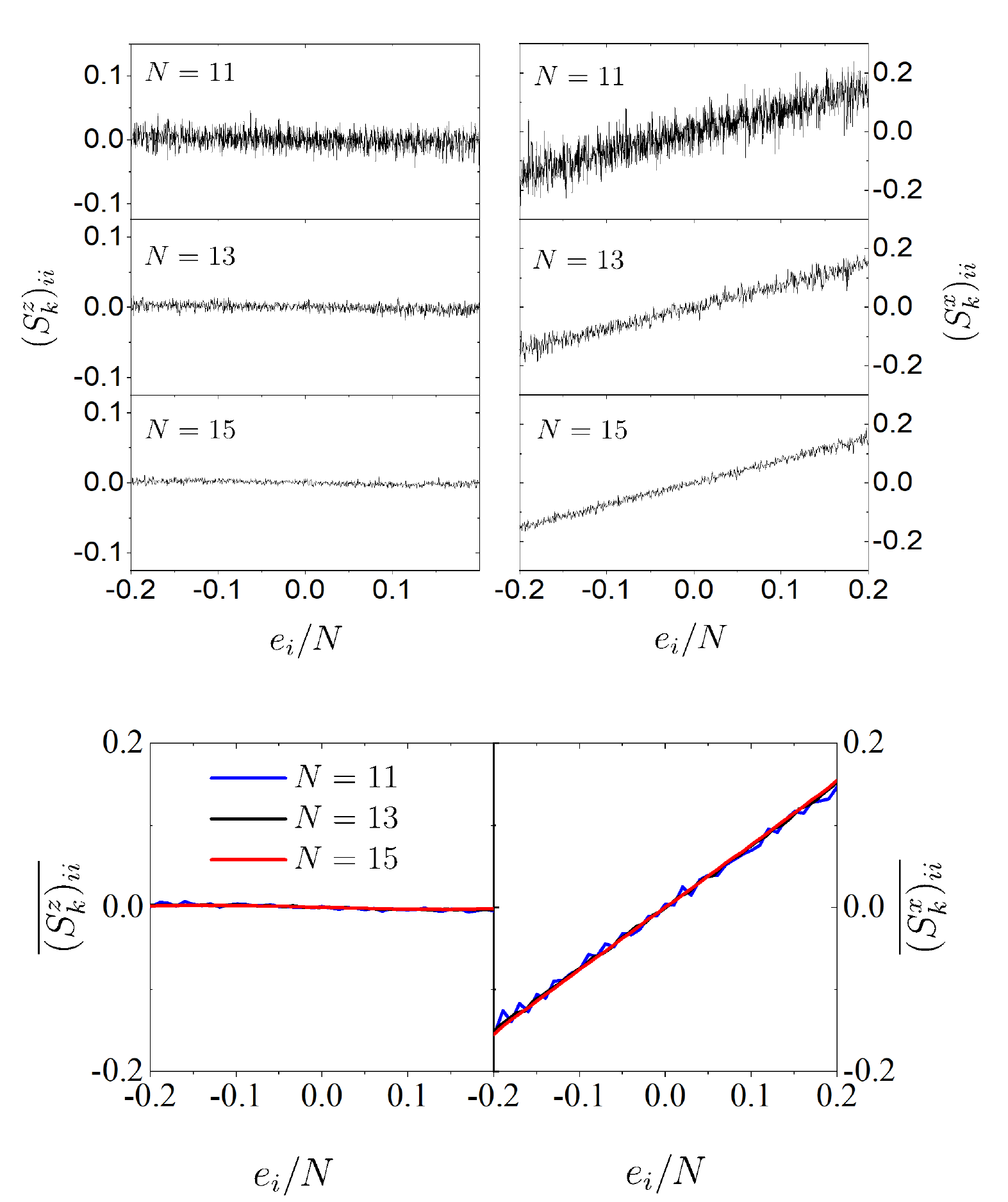}
	\caption{ \label{fig:diag}
	 {Upper panel}: Diagonal matrix elements of two local observables
	 $S_k^z$ and $S_k^x$ ($k=7$) of the defect Ising chain vs the environmental energy
	 $\epsilon_i=e_i/N$ for different chain size $N$. In agreement with the ETH ansatz in Eq.\eqref{ETH},
	 these elements fluctuate around certain slowly-varying functions of $e$,
	 respectively, and the fluctuations decrease with the increase of the particle number $N$.
	 {Lower panel}: Locally averaged values of the above elements (within windows with a width $0.01$),
	 showing a feature of approximate size-independence.
	 }
	\end{figure}

 Due to the hypothesis feature of the ETH ansatz in Eq.\eqref{ETH},
 we have checked its validity in the model employed in this paper. We did this for the two local operators
 $S_k^x$ and $S_k^z$ at the site $k=7$ in the defect Ising chain.
 \begin{figure}
	\includegraphics[width=1.0\linewidth]{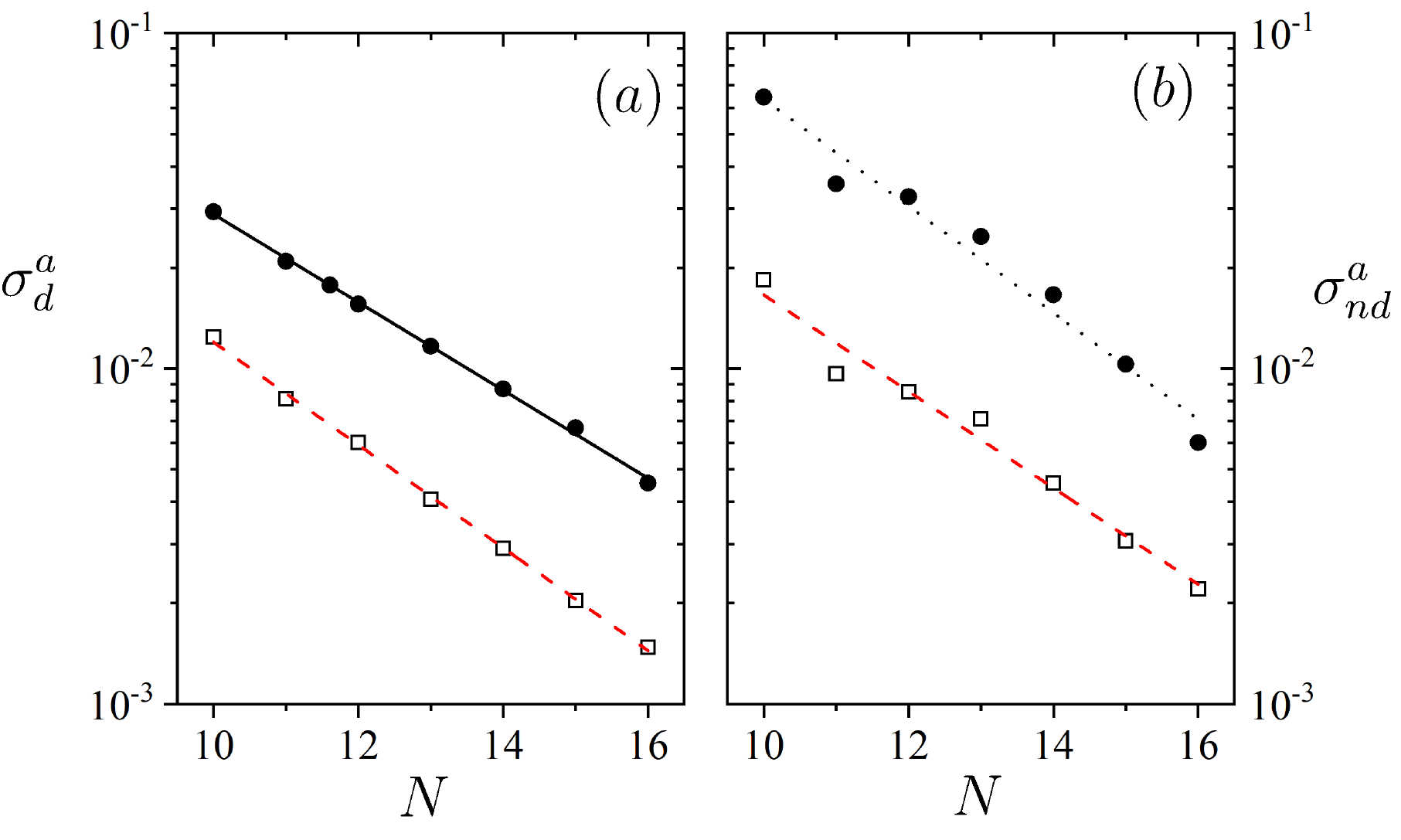}
	\caption{\label{fig:fluc} (a) Exponential decay of the deviation $\sigma_d^a$ in Eq.\eqref{sigma-da}
		with the increase of $N$,
		for fluctuations of the diagonal elements of $S_k^x$ (empty squares) and of $S_k^z$ (solid circles).
		(b) Exponential decay of $\sigma^a_{nd}$ in Eq.\eqref{sigma-nda} for fluctuations of offdiagonal elements.
		The results are in agreement with the prediction of ETH in Eq.\eqref{ETH}.
}
\end{figure}

 \emph{Diagonal ETH}.---
 Let us first discuss predictions of Eq.\eqref{ETH} for diagonal elements of local observables.
 Expectation values of the two local observables,
\begin{gather}\label{Skx-ij}
 (S_k^a)_{ii} = \la i |S_k^a|i\ra \quad \text{with $a=x,z$,}
\end{gather}
 are plotted  in Fig.\ref{fig:diag}.
 It is seen that, in agreement with ETH, the diagonal elements fluctuate around
 certain slowly varying function $h(e)$ and the fluctuations decrease with increasing the particle number $N$.
 Note that the horizontal axis is labeled by $e_i/N$.
 For $a=z$, the values of $h(e)$ are close to zero, while, for $a=x$, most of $|h(e)|$ are notably larger than zero.

\begin{figure}
	\includegraphics[width=1.0\linewidth]{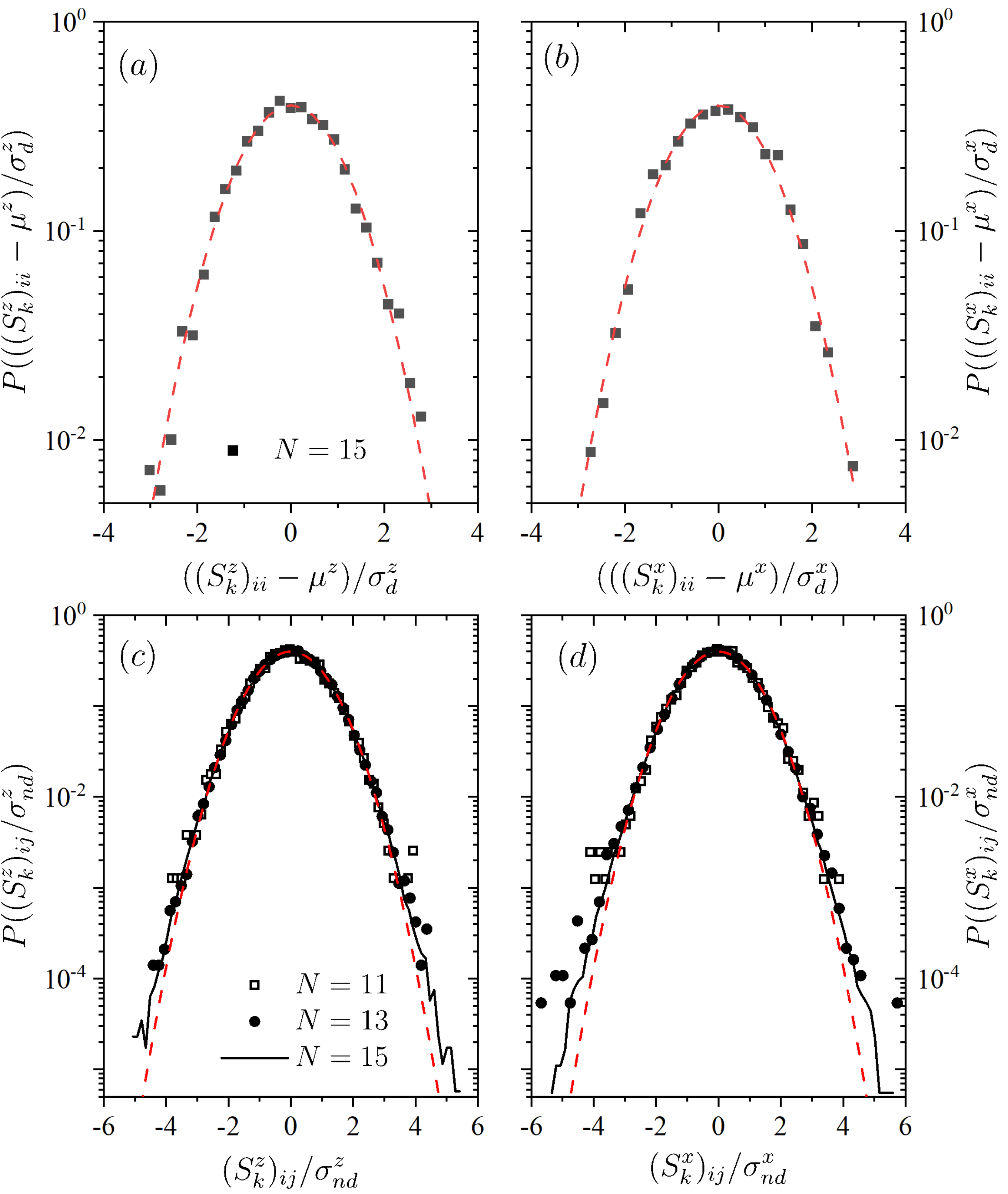}
	 \caption{\label{fig:nondiag} Distributions of fluctuations of the diagonal elements of  $S_k^z$
		 (a) and of $S_k^x$ (b), rescaled by $\sigma_{d}^z$ and $\sigma_{d}^x$, respectively.
	 And, distributions of the offdiagonal
	 elements of $S_k^z$ (c) and of $S_k^x$ (d), rescaled by $\sigma_{nd}^z$ and $\sigma_{nd}^x$, respectively.
	 The dashed curves represent the Gaussian distribution with unit variance.
	 }
	\end{figure}

 To study quantitatively the fluctuations of $(S_k^a)_{ii}$,
 we have computed the standard deviations  $\sigma^a_d$,
 \begin{equation}\label{sigma-da}
\sigma_d^a = \sqrt{\frac{1}{N_{\Gamma_0^\E}}\sum_{e_i\in\Gamma_0^\E}|(S_k^a)_{ii}-\mu^a|^2},
\end{equation}
where
\begin{equation}
\mu^a = \frac{1}{N_{\Gamma_0^\E}}\sum_{e_i\in\Gamma_0^\E} (S_k^a)_{ii}.
\end{equation}
 As seen in Fig.~\ref{fig:fluc} (a), the fluctuation decays exponentially with the increase of $N$,
 as predicted by the term $e^{-S(e)}$ in the second part on the rhs of Eq.\eqref{ETH}.
 Moreover, in agreement with the prediction of ETH, the distributions of $[(S_k^a)_{ii} - \mu^a]/\sigma^a_{d}$
 are close to the Gaussian form [Fig.~\ref{fig:nondiag} (a) and (b)].

	\begin{figure}[ht]
		\centering
		\includegraphics[width=1.0\linewidth]{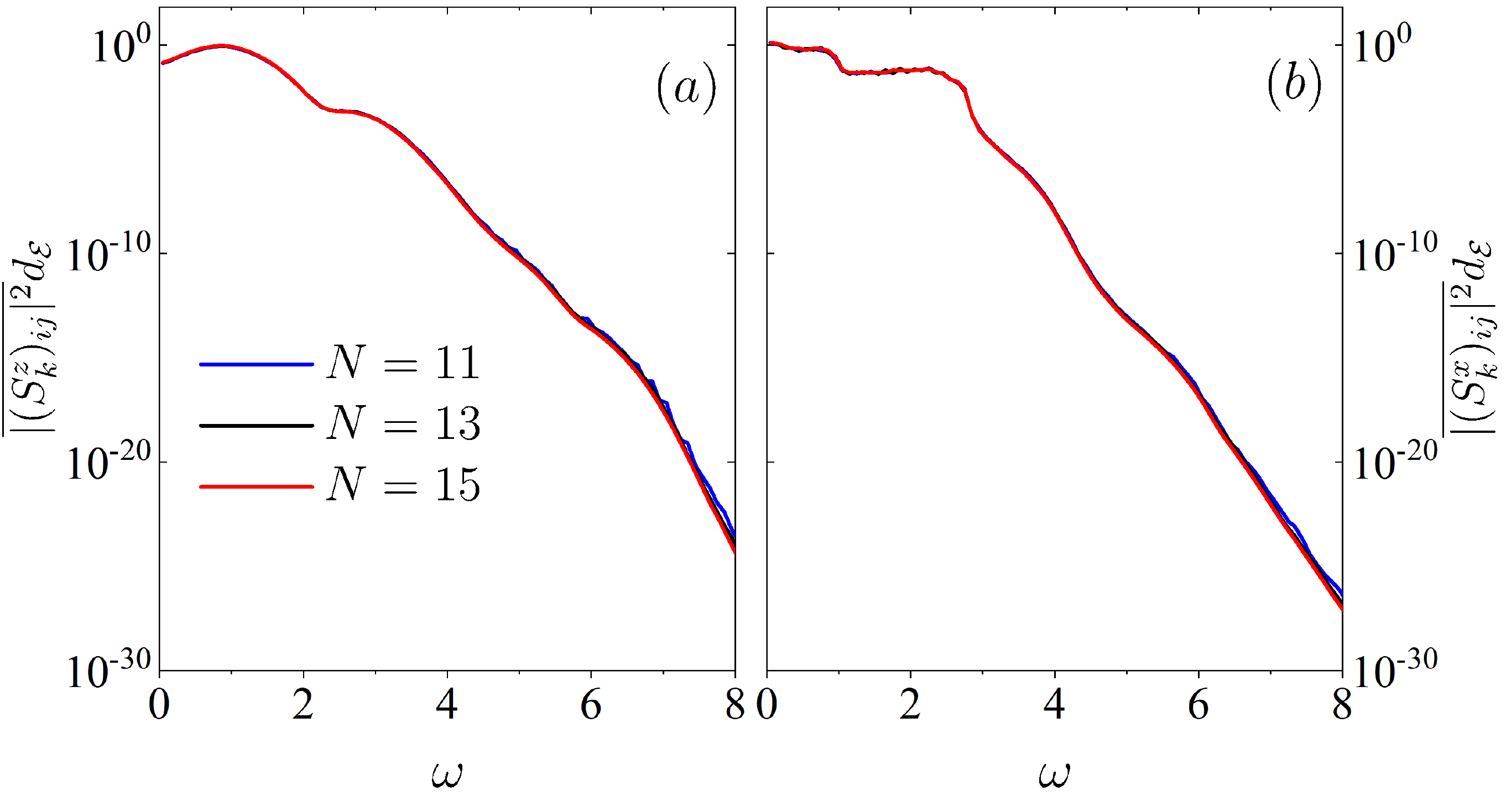}
		\caption{Locally averaged values (in the logarithm scale) of the absolute square of off-diagonal elements of (a) $S_k^z$ and (b) $S_k^x$,
	 within an energy shell centered at $-1.2$ and with a width $0.2$, vs $\omega = e_j-e_i$.
	 Local averages were taken within small windows with width $0.01$.
	}
		\label{fig:formfunc}
	\end{figure}

 \emph{Off-diagonal ETH}.--- Next, we discuss the offdiagonal elements $(S_k^a)_{ij}$.
 In agreement with the prediction of ETH, the probability distributions of $(S_k^a)_{ij}/\sigma^a_{nd}$
 have a Gaussian form [Fig.~\ref{fig:nondiag} (c) and (d)], where $\sigma^a_{nd}$ are the standard deviations
 for the offdiagonal elements,
 \begin{equation}\label{sigma-nda}
\sigma_{nd}^a = \sqrt{\frac{1}{N_{\Gamma_0^\E}(N_{\Gamma_0^\E}-1)}\sum_{i\ne j\in\Gamma_0^\E}|(S_k^a)_{ij}|^2}.
\end{equation}
 These standard deviations also decay exponentially with the increase of $N$ [Fig.~\ref{fig:fluc} (b)].

 To get some knowledge about shapes of the function $g(e,\omega)$, which lacks an analytical expression,
 numerical simulations have been performed for
 locally averaged values of  $\overline{|(S_k^z)_{ij}|^2}$ and $\overline{|(S_k^x)_{ij}|^2}$ for off-diagonal elements.
 As seen in Fig.\ref{fig:formfunc}, the function shows a size-independent feature,
 with an exponential-type decay at large $\omega$.

\bibliography{ssc-cons}

\end{document}